\documentclass[11pt,a4paper]{article}
\pdfoutput=1
\usepackage{jcappub}

\newcommand{\hf}{\frac{1}{2}}
\newcommand{\bea}[1]{\begin{eqnarray} \mbox{$\label{#1}$}}
\newcommand{\eea}{\end{eqnarray}}
\newcommand{\be}[1]{\begin{equation} \mbox{$\label{#1}$}}
\newcommand{\ee}{\vspace{0.1cm}\end{equation}}
\newcommand{\eq}[1]{\mbox{(\ref{#1})}}

\def\mso{m_{s_{0}}}
\newcommand{\lh}{\lambda_h}
\newcommand{\ls}{\lambda_s}
\newcommand{\lhs}{\lambda_{hs}}
\newcommand{\tdN}{\tilde{N}}

\title{WIMP Dark Matter and Unitarity-Conserving Inflation via a Gauge Singlet Scalar}
\author[a]{Felix Kahlhoefer}
\emailAdd{felix.kahlhoefer@desy.de}
\author[b]{and John McDonald}
\emailAdd{j.mcdonald@lancaster.ac.uk}

\affiliation[a]{DESY, Notkestrasse 85, D-22607 Hamburg, Germany}
\affiliation[b]{Lancaster University, Lancaster LA1 4YB, United Kingdom}

\keywords{Dark matter theory, Cosmology of theories beyond the SM, Inflation}

\abstract{A gauge singlet scalar with non-minimal coupling to gravity can drive inflation and later freeze out to become cold dark matter. We explore this idea by revisiting inflation in the singlet direction (S-inflation) and Higgs Portal Dark Matter in light of the Higgs discovery, limits from LUX and observations by Planck. We show that large regions of parameter space remain viable, so that successful inflation is possible and the dark matter relic abundance can be reproduced. Moreover, the scalar singlet can stabilise the electroweak vacuum and at the same time overcome the problem of unitarity-violation during inflation encountered by Higgs Inflation, provided the singlet is a real scalar. The 2-$\sigma$ Planck upper bound on $n_{s}$ imposes that the singlet mass is below 2 TeV, so that almost the entire allowed parameter range can be probed by XENON1T.}

\begin{document}

\begin{flushright}
DESY-15-117
\end{flushright}
\vspace{-9mm}

\maketitle

\section{Introduction}\label{intro}

There is strong evidence in support of the idea that the Universe underwent a period of primordial inflation. In particular, the observation of adiabatic density perturbations with a spectral index which deviates from unity by a few percent~\cite{Ade:2015lrj} is consistent with the generic prediction of scalar field inflation models. However, the identity of the scalar field responsible for inflation remains unknown. Another unsolved problem of similar importance for cosmology is the nature of dark matter (DM). While it is possible to explain DM by the addition of a new particle, there is presently no experimental evidence for its existence or its identity.

The most studied proposal is that DM is a thermal relic weakly-interacting massive particle (WIMP). WIMPs typically have annihilation cross-sections comparable to the value required to reproduce the observed density of DM, the so-called ``WIMP miracle''.  Nevertheless, the non-observation of any new weak-scale particles at the LHC beyond the Standard Model (SM) places strong constraints on many models for WIMPs, such as in supersymmetric extensions of the SM. The absence of new particles may indeed indicate that any extension of the SM to include WIMP DM should be rather minimal. In the present work we therefore focus on a particularly simple extension of the SM, namely an additional gauge singlet scalar, which is arguably one of the most minimal models of DM~\cite{Silveira:1985rk,McDonald:1993ex,Burgess:2000yq}.

A similar issue arises from recent constraints on inflation. In fact, the non-observation of non-Gaussianity by Planck~\cite{Ade:2015ava} suggests that the inflation model should also be minimal, in the sense of being due to a single scalar field. The absence of evidence for new physics then raises the question of whether the inflaton scalar can be part of the SM or a minimal extension of the SM. The former possibility is realized by Higgs Inflation~\cite{Bezrukov:2007ep}, which is a version of the non-minimally coupled scalar field inflation model of Salopek, Bond and Bardeen (SBB)~\cite{Salopek:1988qh} with the scalar field identified with the Higgs boson. A good example for the latter option are gauge singlet scalar extensions of the SM, because the DM particle can also provide a well-motivated candidate for the scalar of the SBB model. In other words, in these models the same scalar particle drives inflation and later freezes out to become cold DM.

The resulting gauge singlet inflation model was first considered in~\cite{Lerner:2009xg}, where it was called S-inflation (see also~\cite{Okada:2010jd}).\footnote{The case of singlet DM added to Higgs Inflation was considered in~\cite{Clark:2009dc}.} All non-minimally coupled scalar field inflation models based on the SBB model are identical at the classical level but differ once quantum corrections to the inflaton potential are included. These result in characteristic deviations of the spectral index from its classical value, which have been extensively studied in both Higgs Inflation~\cite{Bezrukov:2007ep, Bezrukov:2008ej, DeSimone:2008ei, Barvinsky:2008ia, Bezrukov:2009db, Barvinsky:2009fy} and S-inflation~\cite{Lerner:2011ge}.   

Since the original studies were performed, the mass of the Higgs boson~\cite{Aad:2015zhl} and the Planck results for the inflation observables~\cite{Ade:2015lrj} have become known. In addition, direct DM detection experiments, such as LUX~\cite{Akerib:2013tjd}, have imposed stronger bounds on gauge singlet scalar DM~\cite{Cline:2013gha,Mambrini:2011ik,Djouadi:2011aa, Djouadi:2012zc,deSimone:2014pda,Feng:2014vea,Queiroz:2014yna}. This new data has important implications for these models, in particular for S-inflation, which can be tested in Higgs physics and DM searches. The main objective of the present paper is to compare the S-inflation model with the latest results from CMB observations and direct DM detection experiments.  

We will demonstrate that~--- in spite of its simplicity~--- the model still has a large viable parameter space, where the predictions for inflation are consistent with all current constraints and the observed DM relic abundance can be reproduced. In addition, we observe that this model can solve the potential problem that the electroweak vacuum may be metastable, because the singlet gives a positive contribution to the running of the quartic Higgs coupling. Intriguingly, the relevant parameter range can be almost completely tested by XENON1T. 

Another important aspect of our study is perturbative unitarity-violation, which may be a significant problem for Higgs Inflation. Since Higgs boson scattering via graviton exchange violates unitarity at high energies~\cite{Barbon:2009ya,Burgess:2009ea}, one might be worried that the theory is either incomplete or that perturbation theory breaks down so that unitarity is only conserved non-perturbatively~\cite{Han:2004wt,Lerner:2010mq,Lerner:2011it,Aydemir:2012nz}. In both cases there can be important modification of the inflaton potential due to new physics or strong-coupling effects. Indeed, in conventional Higgs Inflation, the unitarity-violation scale is of the same magnitude as the Higgs field during inflation~\cite{Bezrukov:2009db,Bezrukov:2010jz}, placing in doubt the predictions of the model or even its viability. 

In contrast, we will show that S-inflation has sufficient freedom to evade this problem, provided that the DM scalar is specifically a {\it real} singlet. By choosing suitable values for the non-minimal couplings at the Planck scale, it is possible for the unitarity-violation scale to be much larger than the inflaton field throughout inflation, so that the predictions of the model are robust. Therefore, in addition to providing a minimal candidate for WIMP DM, the extension of the SM by a non-minimally coupled real gauge singlet scalar can also account for inflation while having a consistent scale of unitarity-violation.

The paper is organized as follows. In section~\ref{model} we review the real gauge singlet scalar extension of the SM and the S-inflation model. We estimate the predictions of the model for the spectral index $n_{s}$ and discuss the effect of constraints from inflation on the model parameter space. Section~\ref{dm} considers the DM phenomenology of the model and the implications from DM searches. 
In section~\ref{RG} we discuss how to connect these two aspects via renormalisation group evolution and which constraints follow from electroweak vacuum stability and perturbativity. 
The scale of unitarity-violation during inflation and the consistency of S-inflation are discussed in section~\ref{unitarity}.  Finally, we present our results in section~\ref{results} and our conclusions in section~\ref{conclusions}. Additional details are provided in the Appendix. 

\section{The S-inflation model}\label{model}

S-inflation is a version of the non-minimally coupled inflation model of~\cite{Salopek:1988qh} in which the scalar field is identified with the gauge singlet scalar responsible for thermal relic cold DM. In the present work, we focus on the case of a real singlet scalar $s$. In the Jordan frame, which is the standard frame for interpreting measurements and calculating radiative corrections, the action for this model is
\bea{e1}
 S_\text{J} & = & \int \sqrt{-g} \, \mathrm{d}^4 x \Big[ {\cal L}_{\overline{\text{SM}}} + \left(\partial _\mu H\right)^{\dagger}\left(\partial^\mu H\right) + \frac{1}{2} \partial _\mu s \, \partial^\mu s \nonumber \\
 & &  - \frac{m_\text{P}^2 \, R}{2} - \xi_h \, H^{\dagger}H \, R - \frac{1}{2} \, \xi_s \, s^2 \, R - V(s^2,H^{\dagger}H)\Big]
\;,\eea
where ${\cal L}_{\overline{\text{SM}}}$ is the SM Lagrangian density minus the purely Higgs doublet terms, $m_\text{P}$ is the reduced Planck mass and
\bea{e2}
V(s^2,H^{\dagger}H) = \lh \left[\left(H^{\dagger}H\right) - \frac{v^2}{2}\right]^2 + \frac{1}{2} \, \lhs \, s^2 \, H^{\dagger}H + \frac{1}{4} \, \ls \, s^4 + \frac{1}{2} \, \mso^2 \, s^2\eea
with $v = 246 \: \text{GeV}$ the vacuum expectation value of the Higgs field. Writing $H = (h + v, 0)/\sqrt{2}$ with a real scalar $h$ we obtain
\be{potential}
 V(s^2,h) = V(h) + \frac{1}{2} \, m_{s}^2 \, s^2 + \frac{1}{4} \, \ls \, s^4 + \frac{1}{2} \, \lambda_{hs} \, v \, h \, s^2 + \frac{1}{4} \, \lambda_{hs} \, h^2 \, s^2 \; ,
\ee
where we have introduced the physical singlet mass $m_{s}^2 = \mso^2 + \lambda_{hs} \, v^2 / 2$.

In order to calculate the observables predicted by inflation, we perform a conformal transformation to the Einstein frame, where the non-minimal coupling to gravity disappears.
In the case that $s \neq 0$ and $h = 0$,  this transformation is defined by
\be{e3} \tilde{g}_{\mu\nu} = \Omega ^2 \, g_{\mu\nu} \, , \qquad \Omega^2 = 1 + \frac{\xi_s \, s^2}{m_\text{P}^2} \;.\ee
The transformation yields
\be{e4}
S_\text{E}  =  \int \sqrt{-\tilde{g}} \, \mathrm{d}^4x \left[  \tilde{{\cal L}}_{\overline{SM}} + \frac{1}{2}\left(\frac{1}{\Omega^2} + \frac{6 \, \xi_s^2 \, s^2}{m_\text{P}^2 \, \Omega^4}\right)\tilde{g}^{\mu\nu} \partial_\mu s \partial_\nu s  
-\frac{m_\text{P}^2 \, \tilde{R}}{2}- \frac{V(s,0)}{\Omega^4}\right]
\;,\ee
where $\tilde{R}$ is the Ricci scalar with respect to $\tilde{g}_{\mu \nu}$. We can then rescale the field using 
\be{e5}
\frac{\mathrm{d}\chi_s}{\mathrm{d}s} = \sqrt{\frac{\Omega^2 + 6 \, \xi_s^2 \, s^2/m_\text{P}^2}{\Omega ^4}}
\;,\ee
which gives
\be{e6}
S_\text{E} =  \int \sqrt{-\tilde{g}} \, \mathrm{d}^4x \left( \tilde{{\cal L}}_{\overline{SM}} - \frac{m_\text{P}^2 \, \tilde{R}}{2}
 + \frac{1}{2}\tilde{g}^{\mu\nu}\partial _\mu \chi_s \partial_\nu \chi_s - U(\chi_s,0)\right)
\;,\ee
with
\be{e7} U(\chi_s,0) = \frac{\lambda_{s} \, s^4(\chi_{s})}{4 \, \Omega^4}\;.\ee
The relationship between $s$ and $\chi_{s}$ is determined by the solution to eq.~\eq{e5}. In particular, for $s \gg m_\text{P}/\sqrt{\xi_{s}}$, the Einstein frame potential is 
\be{e8}  U(\chi_s,0) = \frac{\lambda_{s} \, m_\text{P}^{4}}{4 \, \xi_s^2} \left( 1 + \exp\left(-\frac{2 \, \chi_s}{\sqrt{6} \, m_\text{P}}
\right)\right)^{-2} \;.\ee
This is sufficiently flat at large $\chi_{s}$ to support slow-roll inflation. 

An analogous expression is obtained for the potential along the $h$-direction. In both cases, the Einstein frame potential is proportional to 
$\lambda_\phi / \xi_{\phi}^{2}$, where $\phi = s$ or $h$. 
Therefore the minimum of the potential at large $s$ and $h$ will be very close to $h = 0$ and inflation will naturally occur along the $s$-direction if $\ls/\xi_s^2 \ll \lh/\xi_h^2$, which is true for example if $\xi_{s} \gg \xi_{h}$ and $\lambda_{s} \sim \lambda_{h}$. 

In the following, inflation is always considered to be in the direction of $s$ with $h = 0$. The conventional analysis of inflation can then be performed in the Einstein frame. After inflation, the Jordan and Einstein frames will be indistinguishable since $\xi_{s} \, s^{2} \ll m_\text{P}^{2}$ and so $\Omega \rightarrow 1$. Therefore the curvature perturbation spectrum calculated in the Einstein frame becomes equal to that observed in the physical Jordan frame at late times. 

The classical (tree-level) predictions for the spectral index and tensor-to-scalar ratio are~\cite{Lerner:2009na}
\be{e9}
n_s^\text{tree} \approx 1 -\frac{2}{\tdN} - \frac{3}{2 \tdN^2} + {\cal O}\left(\frac{1}{\tdN^3}\right) = 0.965 \;,\ee
\be{e10} r^\text{tree} \approx \frac{12}{\tdN^2}  + {\cal O}\left(\frac{1}{\xi_s \tdN^2}\right) = 3.6\times 10^{-3}  \;,\ee
while the field during inflation is 
\be{e11} s_{\tilde{N}}^2 \approx 4 \, m_\text{P}^2 \, \tilde{N}/ (3 \, \xi_s) \;.\ee  
In the equations above $\tdN$ is the number of e-foldings as defined in the Einstein frame, which differs from that in the Jordan frame by $\tilde{N} \approx N + \ln(1/\sqrt{N})$~\cite{Lerner:2009xg}, and we have used $\tdN = 58$.
\footnote{Reheating in S-inflation occurs via stochastic resonance to Higgs bosons through the coupling $\lambda_{hs}$. It was shown in~\cite{Lerner:2011ge} that this process is very efficient and makes quite precise predictions for the reheating temperature and the number of e-foldings of inflation, with $57 
\lesssim \tilde{N} \lesssim 60$ at the WMAP pivot scale. This in turn allows for quite precise 
predictions of the inflation observables.}
The classical predictions are in good agreement with the most recent Planck values, $n_{s} = 0.9677 \pm 0.0060$ (68$\%$ confidence level (CL), Planck TT + lowP + lensing) and $r_{0.002} < 0.11$ (95$\%$ CL, Planck TT + lowP + lensing)~\cite{Ade:2015lrj}. 

The classical predictions for S-inflation are the same as those of any model based on the SBB model. Differences between S-inflation and other models do however arise from quantum corrections to the effective potential. To include these corrections, we calculate the RG evolution of the various couplings as a function of the renormalisation scale $\mu$ (see section~\ref{RG} for details). We can then obtain the renormalisation group (RG)-improved effective potential for $s$ in the Jordan frame by replacing the couplings in eq.~\eq{potential} by the running couplings and setting $\mu$ equal to the value of the field. For $h=0$ (and neglecting the singlet mass term, $m_s \sim 1\:\text{TeV} \ll s$), this approach yields
\bea{e12}
V_\text{RG}(s^2, 0)  =  \frac{\ls(s) \, s^4}{4}\;.
\eea

The RG-improved potential can then be transformed into the Einstein frame in order to calculate the observables predicted by inflation.\footnote{In~\cite{George:2013iia} it was proposed to use of the Einstein frame for the computation of quantum corrections. The Jordan frame analysis is, however, easier to implement correctly, being a straightforward extension of the Standard Model analysis.}
The inflationary parameters are calculated using the methods discussed in~\cite{Lerner:2011ge}. In particular, the Einstein frame slow-roll parameters are given by
\begin{align}
\tilde{\epsilon} & = \frac{m_\text{P}^2}{2} \left(\frac{1}{U} \frac{\mathrm{d}U}{\mathrm{d}\chi_s}\right)^2 \;, \nonumber \\
\tilde{\eta} & = \frac{m_\text{P}^2}{U}\frac{\mathrm{d}^2U}{\mathrm{d}\chi_s^2} \nonumber \;,\\
\tilde{\xi}^2 & = \frac{m_\text{P}^4}{U^2}\frac{\mathrm{d}U}{\mathrm{d}\chi_s}\frac{\mathrm{d}^3U}{\mathrm{d}\chi_s^3} \; .
\label{eq:slowroll}
\end{align}

\section{Singlet scalar as dark matter}
\label{dm}

Let us now turn to the phenomenology of the singlet scalar in the present Universe and at energies well below the scale of inflation~\cite{Barger:2007im, Cline:2013gha, Feng:2014vea}. Most importantly, the assumed $\mathbb{Z}_2$ symmetry ensures the stability of the scalar, so that it can potentially account for the observed abundance of DM~\cite{Silveira:1985rk,McDonald:1993ex}. If the mass of the singlet is comparable to the electroweak scale, the singlet is a typical WIMP, which obtains its relic abundance from thermal freeze-out. Indeed, at low energies, where the effects of the non-minimal coupling to gravity are negligible, our model becomes identical to what is often referred to as \emph{Higgs Portal Dark Matter}~\cite{Djouadi:2011aa, Djouadi:2012zc, Khoze:2013uia}, because all interactions of the singlet with SM particles are mediated by the Higgs. In this section we review the constraints on these models and determine the parameter space allowed by the most recent experimental results. In the process, we point out several discrepancies in the literature and resolve the resulting confusion. 

\subsection{Relic abundance}

The calculation of the relic abundance of singlet scalars is discussed in detail in~\cite{Cline:2013gha}. Three kinds of processes are relevant for the annihilation of singlets into SM states: annihilation into SM fermions, annihilation into SM gauge bosons and annihilation into two Higgs particles. The first kind dominates as long as $m_s < m_W$, while for larger masses the second kind gives the largest contribution. Notably, all of these processes can proceed via an $s$-channel Higgs boson, leading to a resonant enhancement of the annihilation cross-section and a corresponding suppression of the DM relic abundance for $m_s \sim m_h / 2$.\footnote{The process $ss \rightarrow hh$ also receives a contribution from $t$-channel singlet exchange, which gives a relevant contribution if $\lambda_{hs}$ is large compared to $\lambda_h$.}

For the present work we calculate the singlet abundance using micrOMEGAs\_3~\cite{Belanger:2013oya}, which numerically solves the Boltzmann equation while calculating the Higgs width in a self-consistent way. It is then straightforward to numerically find the coupling $\lambda_{hs}$ that gives $\Omega_s \, h^2 = 0.1197$, in order to reproduce the value of the DM density $\Omega_\text{DM} \, h^2 = 0.1197 \pm 0.0022$ determined by Planck (TT + lowP, 68$\%$ CL)~\cite{Ade:2015xua}. For example, we find $\lambda_{hs} \approx 0.08$ for $m_s = 300\:\text{GeV}$ and $\lambda_{hs} \approx 0.30$ for $m_s = 1000\:\text{GeV}$. These values agree with the ones found in~\cite{Cline:2013gha,deSimone:2014pda}, but disagree with~\cite{Feng:2014vea, Mambrini:2011ik, Djouadi:2011aa, Djouadi:2012zc} by a factor of 2 after accounting for the different conventions.\footnote{The Higgs-singlet coupling is called $\lambda_{hS}$ in~\cite{Cline:2013gha}, $\lambda_{HS}$ in~\cite{Mambrini:2011ik}, $\lambda_{hSS}$ in~\cite{Djouadi:2011aa, Djouadi:2012zc}, $\lambda_\text{DM}$ in~\cite{deSimone:2014pda} and $a_2$ in~\cite{Feng:2014vea}. The respective conventions are captured by $ \lambda_{hs} = \lambda_{hS} = \lambda_{HS}/2 = \lambda_{hSS}/2 = \lambda_\text{DM}/2 = 2 \, a_2 $.}

\subsection{Direct detection constraints}
 
The strongest constraints on $\lambda_{hs}$ stem from DM direct detection experiments, since the singlet-Higgs coupling induces spin-independent interactions between the DM particle and nuclei. The scattering cross-section at zero momentum transfer is given by~\cite{Cline:2013gha}
\begin{equation}
 \sigma_\text{SI} = \frac{\lambda_{hs}^2 f_N^2}{4\pi} \frac{\mu_\text{r}^2 \, m_n^2}{m_h^4 \, m_s^2} \; ,
\end{equation}
where $m_n$ is the neutron mass, $\mu_\text{r} = (m_s \, m_n) / (m_s + m_n)$ is the reduced mass and $f_N$ is the effective Higgs-nucleon coupling.\footnote{Note that~\cite{Feng:2014vea} uses an approximate expression valid for $m_s \gg m_n$, such that $\mu_\text{r}^2 \approx m_n^2$.} In terms of the light-quark matrix elements $f_{Tq}^N$, the effective coupling can be written as
\begin{equation}
f_N = \left[\frac{2}{9} + \frac{7}{9} \, \sum_{q=u,d,s} f_{Tq}^N \right] \;. 
\end{equation}
The values of $f_{Tq}^N$ can either be determined phenomenologically from baryon masses and meson-baryon scattering data or  computed within lattice QCD. A comparison of the different methods was recently performed in~\cite{Cline:2013gha} and we adopt their result of $f_N = 0.30$ for the effective coupling.

\begin{figure}[tb]
\centering
\includegraphics[width=0.5\textwidth]{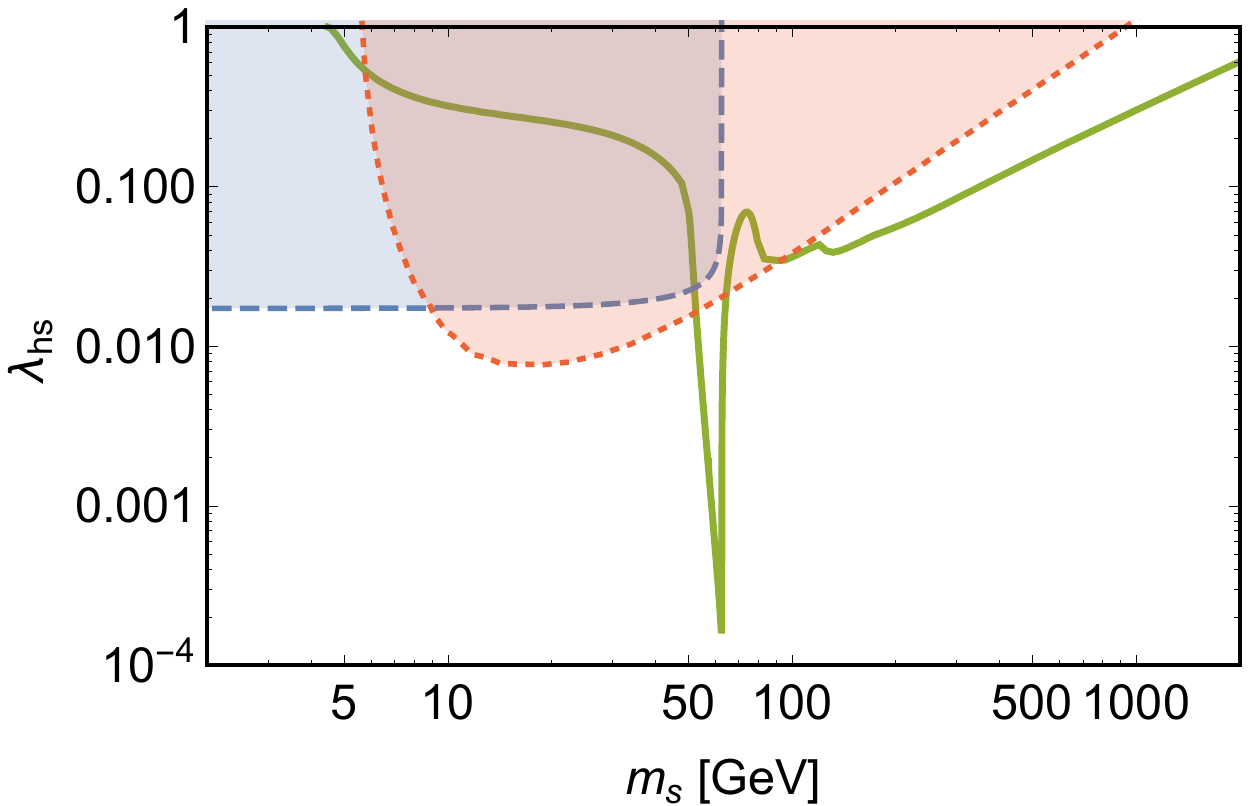}
\caption{Excluded parameter regions from LUX (red, dotted) and searches for invisible Higgs decays (blue, dashed) compared to the coupling implied by the relic density constraint (green, solid).}
\label{fig:DM}
\end{figure}

The scattering cross-section given above can be directly compared to the bound obtained from the LUX experiment~\cite{Akerib:2013tjd}. Indeed, as shown in figure~\ref{fig:DM}, LUX is typically sensitive to the same range of values for $\lambda_{hs}$ as what is implied by the relic density constraint. Specifically, the LUX bound excludes the mass ranges $5.7\:\text{GeV} < m_s < 52.6 \: \text{GeV}$ and $64.5 \: \text{GeV} < m_s < 92.8 \: \text{GeV}$.

\subsection{Invisible Higgs decays}

Direct detection experiments cannot constrain singlet scalars with a mass of a few GeV or less, since such particles would deposit too little energy in the detector to be observable. This parameter region can however be efficiently constrained by considering how the Higgs-singlet coupling $\lambda_{hs}$ would modify the branching ratios of the SM Higgs boson. The partial decay width for $h \rightarrow ss$ is given by\footnote{This equation agrees with~\cite{Burgess:2000yq,Barger:2007im,Cline:2013gha,Mambrini:2011ik, Djouadi:2011aa, Djouadi:2012zc} but disagrees with~\cite{deSimone:2014pda}.}
\begin{equation}
\Gamma(h \rightarrow ss) = \frac{\lambda_{hs}^2 v^2}{32\pi \, m_h} \sqrt{1 - \frac{4 \, m_s^2}{m_h^2}} \; .
\end{equation}

This theoretical prediction can be compared to the experimental bound on invisible Higgs decays from the LHC. Direct searches for invisible Higgs decays in the vector boson fusion channel give $\text{BR}(h \rightarrow \text{inv}) \lesssim 0.29$~\cite{ATLAS}. A somewhat stronger bound can be obtained from the observation that in our model there are no additional contributions to the Higgs production cross-section and no modifications of the partial decay widths of the Higgs boson into SM final states. Therefore the presence of an invisible decay channel leads to an overall reduction of the signal strength in visible channels. A global fit of all observed decay channels (combined with the bounds on invisible Higgs decays) then gives $\text{BR}(h \rightarrow \text{inv}) \lesssim 0.26$~\cite{Khachatryan:2014jba}.

Crucially, the bound from invisible Higgs decays becomes independent of the singlet mass for \mbox{$m_s \ll m_h / 2$}. Invisible Higgs decays will therefore provide the strongest constraints for small singlet masses. Indeed, this constraint rules out the entire mass region where direct detection experiments lose sensitivity (see figure~\ref{fig:DM}). As a result, only two mass regions remain viable: a low-mass region $52.6 \: \text{GeV} < m_s < 64.5 \: \text{GeV}$ and a high-mass region $m_s \gtrsim 93 \: \text{GeV}$.

\subsection{Other constraints}

It has been pointed out recently~\cite{Feng:2014vea} that bounds on $\gamma$-ray lines from Fermi-LAT~\cite{Ackermann:2013uma} rule out the parameter region where $m_s$ is slightly above $m_h / 2$. To be safe from this constraint, we will focus on the mass range $52.6 \: \text{GeV} < m_s < 62.4 \: \text{GeV}$, which we shall refer to as the low-mass region. For the high-mass region, on the other hand, there are no strong constraints from indirect detection. Moreover, collider searches for singlet scalars with $m_s > m_h / 2$ are extremely challenging~\cite{Barger:2007im, Mambrini:2011ik, Djouadi:2011aa, Djouadi:2012zc, deSimone:2014pda} and consequently, there are no relevant bounds from the LHC for the high-mass region~\cite{Craig:2014lda}.

The most significant improvements in sensitivity in the near future are expected to come from direct detection experiments. Indeed, XENON1T~\cite{XENON1T} is expected to improve upon current LUX constraints on the DM scattering cross section by a factor of about 50 and will therefore be able to probe the high-mass region up to $m_s \approx 4 \: \text{TeV}$. As we will show, in S-inflation singlet masses larger than about 2 TeV are excluded by the Planck 2-$\sigma$ upper bound on $n_{s}$ and perturbativity. XENON1T will therefore be able to probe the \emph{entire high-mass region} relevant for singlet inflation. Similarly, XENON1T can also further constrain the low-mass region and potentially probe singlet masses in the range $53 \: \text{GeV} < m_s < 57 \: \text{GeV}$.

\section{Renormalisation group evolution and theoretical constraints}
\label{RG}

In order to connect the inflationary observables of our model to the measured SM parameters and the DM phenomenology discussed in the previous section, we need to calculate the evolution of all couplings under the RG equations~\cite{Espinosa:2007qp,DeSimone:2008ei,Bezrukov:2009db,Clark:2009dc,Lerner:2009xg,EliasMiro:2011aa,Degrassi:2012ry,Allison:2013uaa}. 
Existing analyses have considered the RG equations for the SM at two-loop order and the contributions of the singlet sector and non-minimal coupling at one-loop order (see also the Appendix)~\cite{Clark:2009dc,Lerner:2009xg}. To examine the issue of vacuum stability, we improve the accuracy of our analysis further by incorporating the three-loop RG equations for the SM gauge couplings~\cite{Mihaila:2012fm} and the leading order three-loop corrections to the RG equations for $\lambda_h$ and $y_t$~\cite{Chetyrkin:2012rz}.\footnote{We thank Kyle Allison for sharing his numerical implementation of these equations.}

When considering large field values for either $s$ or $h$, the RG equations are modified, because there is a suppression of scalar propagators. This suppression is captured by inserting a factor
\begin{equation}
 c_\phi = \frac{ 1 + \frac{\xi_\phi \, \phi^2}{m_\text{P}}}{1 + (6\xi_\phi + 1)\frac{\xi_\phi \, \phi^2}{m_\text{P}}}
\end{equation}
with $\phi = s$ ($\phi = h$) for each $s$ ($h$) propagating in a loop~\cite{Lerner:2009xg}. The changes in the RG equations for large values of the Higgs field have been discussed in detail in~\cite{Allison:2013uaa}. The modifications resulting from large singlet field values can be found in~\cite{Lerner:2009xg} and are reviewed in the Appendix. Note that, when considering S-inflation, such that $s \gg h$, we can set the suppression factor $c_h = 1$.

We determine the values of the SM parameters at $\mu = m_t$ following~\cite{Degrassi:2012ry}. Using the most recent values from the Particle Data Group~\cite{Agashe:2014kda}
\begin{equation}
 m_t = (173.2 \pm 0.9)\:\text{GeV}, \quad m_H = (125.09 \pm 0.24)\:\text{GeV}, \quad \alpha_S(m_Z) = 0.1185 \pm 0.0006
\;,\end{equation}
we obtain at $\mu = m_{t}$
\begin{equation}
y_t = 0.936 \pm 0.005, \quad \lambda_h = 0.1260 \pm 0.0014, \quad g_S = 1.164 \pm 0.003 \; .
\end{equation}
Unless explicitly stated otherwise, we will use the central values for all calculations below. For given couplings at $\mu = m_t$, we then use the public code \texttt{RGErun 2.0.7}~\cite{RGErun} to calculate the couplings at higher scales. 

In contrast to the remaining couplings, we fix the non-minimal couplings $\xi_h$ and $\xi_s$ at $\mu = m_\text{P}$. In order to obtain the correct amplitude of the scalar power spectrum, we require
\begin{equation}
\frac{U}{\tilde{\epsilon}} = (0.00271 \, m_\text{P})^4 \; , 
\label{eq:scalaramplitude}
\end{equation}
where $U$ and $\tilde{\epsilon}$ are the potential and the first slow-roll parameter in the Einstein frame at the beginning of inflation, as defined in eq.~(\ref{e7}) and eq.~(\ref{eq:slowroll}) respectively. Imposing equation~(\ref{eq:scalaramplitude}) allows us to determine $\xi_s$ at the scale of inflation once all other parameters have been fixed. Note that, since the value of $s$ at the beginning of inflation also depends on $\xi_s$, equation~(\ref{eq:scalaramplitude}) can only be solved numerically. We then iteratively determine the required value of $\xi_s$ at the electroweak scale such that RG evolution yields the desired value at the scale of inflation.

The coupling $\xi_h$ plays a very limited role for the phenomenology of our model because we do not consider the case of large Higgs field values for inflation. As a result our predictions for the inflationary observables show only a very mild dependence on $\xi_h$, so that $\xi_h$ can essentially be chosen arbitrarily. Nevertheless, it is not possible to simply set this parameter to zero, since radiative corrections induce a mixing between $\xi_s$ and $\xi_h$. Moreover, we will see below that the value of $\xi_h$ plays an important role for determining whether our model violates unitarity below the scale of inflation. As with $\xi_s$, we fix $\xi_h$ at $m_\text{P}$ and then determine iteratively the required value of $\xi_h$ at the electroweak scale.\footnote{The running of $\xi_h$ between $m_\text{P}$ and the scale of inflation is completely negligible, since the relevant diagrams are strongly suppressed for large values of $s$.}

\subsection{Metastability}
\label{theo}

We now discuss various theoretical constraints related to the RG evolution of the parameters in our model.
It is a well-known fact that for the central values of the measured SM parameters, the electroweak vacuum becomes metastable at high scales, because the quartic Higgs coupling $\lambda_h$ runs to negative values (see e.g.~\cite{Degrassi:2012ry}). This metastability is not in any obvious way a problem, as the lifetime of the electroweak vacuum is well above the age of the Universe~\cite{Buttazzo:2013uya} (note, however, that this estimate may potentially be spoiled by effects from Planck-scale higher-dimensional operators~\cite{Branchina:2013jra,Branchina:2014usa,Branchina:2014rva}).
However, one may speculate that a stable electroweak vacuum is necessary for a consistent theory, for example if the vacuum energy relative to the absolute minimum is a physical energy density leading to inflation.
It is therefore an interesting aspect of singlet extensions of the SM that scalar singlets give a positive contribution to the running of $\lambda_h$~\cite{Gonderinger:2009jp, Profumo:2010kp}:
\begin{equation}
 \beta_{\lambda_h} = \beta_{\lambda_h}^\text{SM} + \frac{1}{32 \pi^2} \, c_s^2 \, \lambda_{hs}^2 \; .
\end{equation}
In fact, it was shown in~\cite{Khan:2014kba} that, for the case of a minimally-coupled singlet, $\lambda_{hs}$ can be chosen such that the electroweak vacuum remains stable all the way up to the Planck scale and at the same time (for appropriate choices of the singlet mass $m_s$) the singlet obtains a thermal relic density compatible with the observed DM abundance.\footnote{Note that if the singlet mixes with the Higgs, there will be additional threshold effects at $\mu = m_s$ from integrating out the singlet~\cite{EliasMiro:2012ay}. In the setup we consider, however, this effect is not important~\cite{Lebedev:2012zw}.}

In order to study electroweak vacuum stability, we need to consider the potential in the $h$-direction with $s = 0$.\footnote{A more detailed study of the potential in general directions with both $s\neq 0$ and $h\neq0$ (along the lines of~\cite{Ballesteros:2015iua}) is beyond the scope of the present work.} Vacuum stability then requires that $\lambda_{h}(\mu) > 0$ for $\mu$ up to $m_\text{P}$.  
In the present study, we consider both the case where $\lambda_{hs}$ is sufficiently large to stabilise the electroweak vacuum and the case where $\lambda_{hs}$ only increases the lifetime of the metastable vacuum, but does not render it completely stable. We focus throughout on the case where $\lambda_{hs}$ is positive.

\subsection{Examples}

\begin{figure}[tb]
\centering
\includegraphics[height=0.18\textheight]{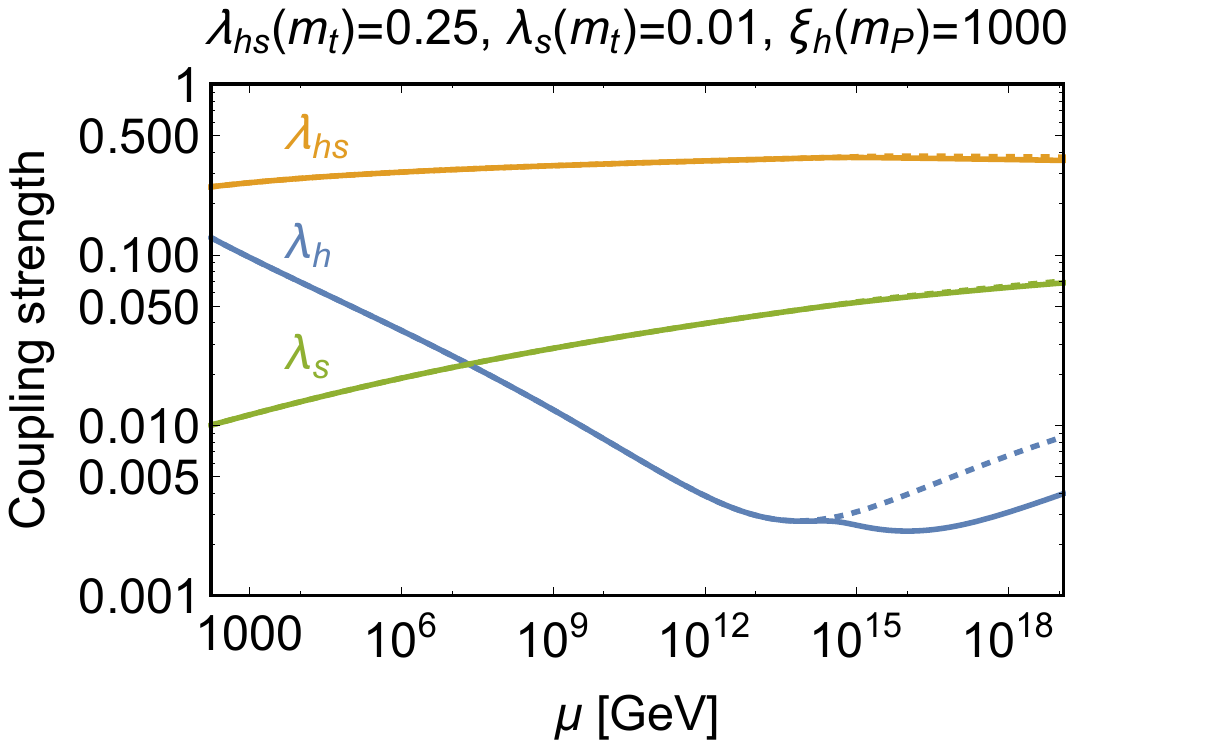}
\includegraphics[height=0.18\textheight]{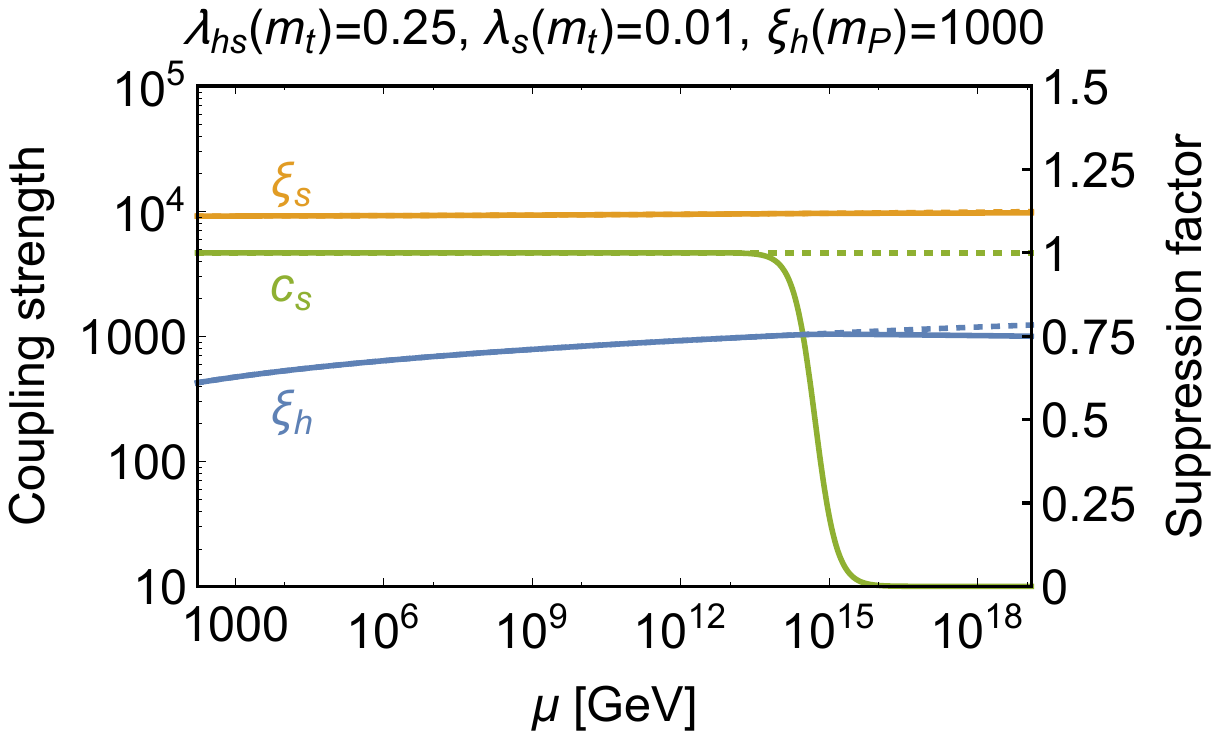}
\caption{Running of the scalar couplings $\lambda_h$, $\lambda_s$ and $\lambda_{hs}$ (left) and of the non-minimal couplings $\xi_h$ and $\xi_s$ (right) as a function of the renormalisation scale $\mu$ for a typical parameter point in the high-mass region. Solid lines show the running in the $s$-direction, while dotted lines correspond to the running in the $h$-direction. In the right panel, we also show the suppression factor $c_s$, which modifies the running in the $s$-direction at large field value.}
\label{fig:running_high}
\end{figure}

Figure~\ref{fig:running_high} shows an example for the evolution of scalar couplings (left) and the non-minimal couplings (right) under the RG equations discussed above. Solid lines correspond to the case $s \gg h$, which is relevant for inflation, while dotted lines correspond to $h \gg s$, which is relevant for vacuum stability. We fix the scalar couplings at the weak scale, choosing $\lambda_{hs} = 0.25$ and $\lambda_{s} = 0.01$, such that the observed relic abundance can be reproduced for $m_s = 835\:\text{GeV}$. We consider $\lambda_{s} \ll \lambda_{hs}$, in which case the value of $\xi_s$ necessary to obtain the correct amplitude of the scalar power spectrum is reduced. For our choice, we find $\xi_s \sim 10^4$. Note, however, that while $\lambda_{hs}$ exhibits only moderate running, $\lambda_{s}$ grows significantly with the renormalisation scale $\mu$, because its $\beta$-function contains a term proportional to $\lambda_{hs}^2$. Choosing even smaller values of $\lambda_s$ at the weak scale will therefore not significantly reduce its value at the scale of inflation nor the corresponding value of $\xi_s$.

An important observation from figure~\ref{fig:running_high} is that $\lambda_h$ does not run negative and hence the electroweak vacuum remains stable all the way up to the Planck scale. The additional contribution from the singlet scalar is sufficient to ensure $\lambda_h > 10^{-3}$ for all renormalisation scales up to $m_\text{P}$. For field values $s \gtrsim 10^{15}\:\text{GeV} \gg h$, the singlet propagator is suppressed, leading to a visible kink in the running of $\lambda_h$. We show the propagator suppression factor $c_s$ in the right panel of figure~\ref{fig:running_high}. One can clearly see how this suppression factor affects the running of $\xi_h$, which becomes nearly constant for $s \gtrsim 10^{15}\:\text{GeV}$. In this particular example, we have chosen $\xi_h(m_\text{P}) = 1000$ (for $s \gg h$). This choice, together with $\lambda_s \ll 1$, implies that the running of $\xi_s(\mu)$ from $m_t$ to $m_\text{P}$ is negligible.

\begin{figure}[tb]
\centering
\includegraphics[width=0.44\textwidth]{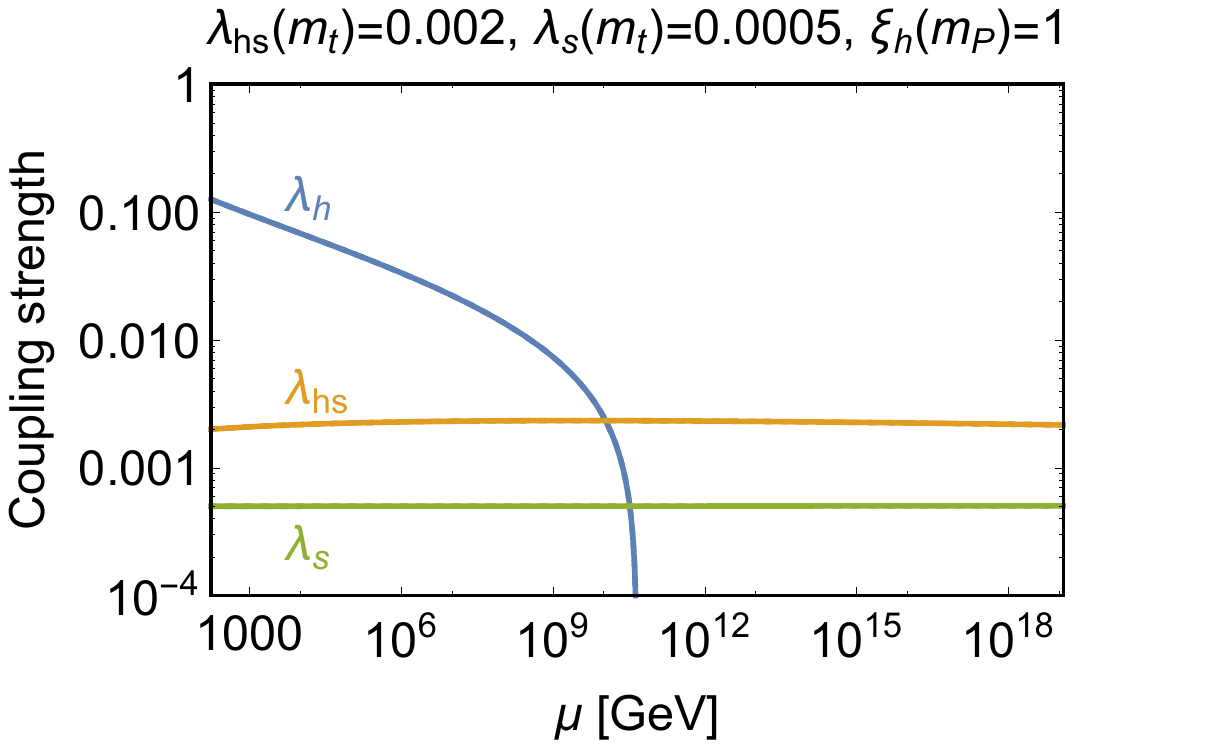}\quad
\includegraphics[width=0.44\textwidth]{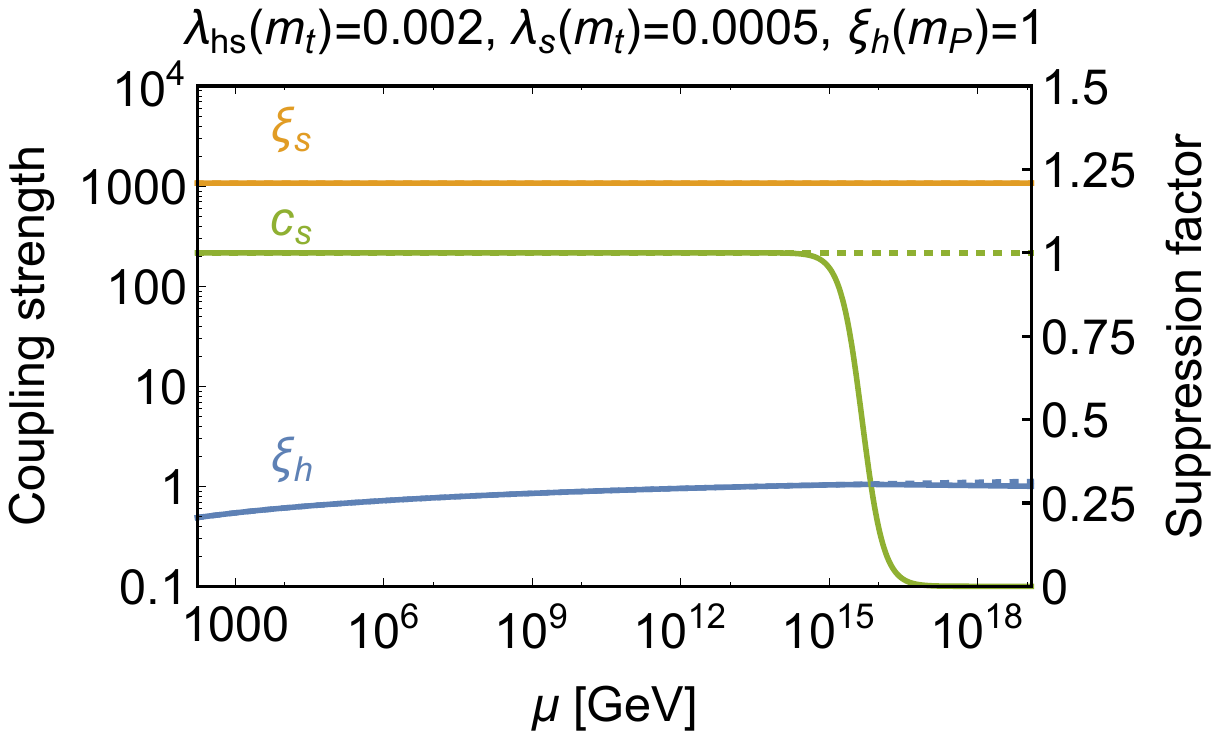}
\caption{Running of the scalar couplings $\lambda_h$, $\lambda_s$ and $\lambda_{hs}$ (left) and of the non-minimal couplings $\xi_h$ and $\xi_s$ (right) as a function of the renormalisation scale $\mu$ for a typical parameter point in the low-mass region. In the right panel, we also show the suppression factor $c_s$, which modifies the running at large field value. Note that $\lambda_h$ runs negative for $\mu \gtrsim 10^{11}\:\text{GeV}$.}
\label{fig:running_low}
\end{figure}

In the low-mass region, we are interested in much smaller values of $\lambda_{hs}$, typically below $10^{-2}$. A particular example is shown in figure~\ref{fig:running_low} (left) for the representative choice $\lambda_{hs}(m_t) = 0.002$ and $\lambda_{s} = 0.0005$, which yields the observed relic abundance for $m_s \approx 57\:\text{GeV}$. We observe that if $\lambda_s$ and $\lambda_{hs}$ are both small at the electroweak scale, these couplings exhibit only very little running up to the scale of inflation. For the same reason, it is impossible to influence the running of the Higgs couplings sufficiently to prevent $\lambda_{hs}$ from running negative at around $10^{11}\:\text{GeV}$.

If $\lambda_{hs}$ is small, we can obtain the correct scalar power spectrum amplitude with a much smaller value of $\xi_s$ during inflation. For the specific case considered in figure~\ref{fig:running_low}, we find $\xi_s \sim 10^3$. For these values of $\xi_s$ and $\lambda_{hs}$, the loop-induced corrections to $\xi_h$ are very small and hence this coupling changes only very slightly under RG evolution.

\subsection{Perturbativity}

In order for our calculation of the running couplings and the radiative corrections to the potential to be reliable, we must require that all couplings remain perturbative up to the scale of inflation, which is typically $10^{17}\text{--}10^{18}\:\text{GeV}$. This requirement is easily satisfied for the SM couplings, but needs to be checked explicitly for the couplings of the singlet, which can grow significantly with increasing renormalisation scale $\mu$. We follow~\cite{Cynolter:2004cq, Khan:2014kba} and use the requirement of perturbative unitarity to impose an upper bound on the scalar couplings. This procedure gives
\begin{equation}
\lambda_s < \frac{4\pi}{3} \qquad \text{and} \qquad \lambda_{hs} < 8 \pi \; .
\end{equation}
As we will see below, the non-minimal coupling $\xi_s$ can be much larger than unity without invalidating a perturbative calculation. Nevertheless, if $\xi_s$ and $\xi_h$ are both very large, processes involving both couplings may violate perturbative unitarity, implying that there may be new physics or strong coupling below the scale of inflation. We will now discuss this issue in more detail.

\section{Unitarity-violation during inflation}

\label{unitarity}

In this section we estimate the scale of perturbative unitarity-violation as a function of the background inflaton field. Note that by ``unitarity-violation scale'' we mean the scale at which perturbation theory in scalar scattering breaks down, so this may in fact indicate the onset of unitarity-conserving scattering in a strongly-coupled regime~\cite{Han:2004wt,Aydemir:2012nz}. We will consider unitarity-violation in the scattering of scalar particles corresponding to perturbations about the background field. In the case of a real scalar $s$ and the fields of the Higgs doublet $H$, there are two distinct scattering processes we need to consider: (i) $\delta s \; h_{1} \leftrightarrow \delta s \; h_{1}$ and (ii)  $h_{1}  h_{2} \leftrightarrow h_{1}  h_{2}$, where 
$\delta s$ is the perturbation about the background $s$ field and $h_{1}$ and $h_{2}$ are two of the Higgs doublet scalars. Scattering with the other scalars in $H$ is equivalent to these two processes.  We will use dimensional analysis to estimate the scale of unitarity-violation by determining the leading-order processes in the Einstein frame which result in unitarity-violating scattering. 

It will be sufficient to consider the Einstein frame Lagrangian for two real scalar fields $\phi_{i}$, where~--- using the notation of~\cite{Lerner:2011it}~--- $\phi_i$ stands for either $s$ or a component of the Higgs doublet. Unitarity-violation requires that there are two different scalars in the scattering process, since in the case of a single scalar there is a cancellation between $s$-, $t$- and $u$-channel amplitudes~\cite{Hertzberg:2010dc}. 

Since unitarity-violating scattering in the Jordan frame is due to graviton exchange via the non-minimal coupling to $R$, we cat set $V = 0$. The Einstein frame action for two real scalars is then of the form
\bea{u1}
S_\text{E} = \int \mathrm{d}^4x\sqrt{-\tilde{g}} \left[{\cal L}_{ii} + \sum_{i < j} {\cal L}_{ij}  -\hf m_\text{P}^2 \, \tilde{R} \right],
\eea
where
\be{u2}
 {\cal L}_{ii} = \hf \left(\frac{\Omega^2+\frac{6 \, \xi_i^2 \, \phi_i^2}{m_\text{P}^2}}{\Omega^4}\right) \tilde{g}^{\mu\nu} \, \partial_\mu \phi_i \, \partial_\nu \phi_i
\quad\text{
and
}
\quad
{\cal L}_{ij} = \frac{6 \, \xi_i \, \xi_j\,\phi_i\,\phi_j \, \tilde{g}^{\mu\nu}\,\partial_\mu \phi_i \, \partial_\nu \phi_j}{m_\text{P}^2 \, \Omega^4}
\ee
with
\be{u3a} \Omega^2 = 1 + \frac{\xi_{j} \, \phi_{j}^{2}}{m_\text{P}^{2}}   \;.\ee
The interaction terms proportional to $\xi_{i} \, \xi_{j}$ are 
responsible for the dominant unitarity-violation in scattering cross-sections calculated in the Einstein frame. These interactions are the Einstein frame analogue of scalar scattering via graviton exchange in the Jordan frame due to the non-minimal coupling. To obtain the scale of unitarity-violation in terms of the physical energy defined in the Jordan frame, we first canonically normalize the fields in the Einstein frame, then estimate the magnitude of the scattering matrix element and finally transform the unitarity-violation scale in the Einstein frame back to that in the Jordan frame. 

In the following we will denote the inflaton by $\phi_{1} \; (\equiv s$), which we expand about the background field, i.e.\ $\phi_{1} = \overline{\phi}_{1} + \delta \phi_{1}$. The Higgs doublet scalars are denoted by $\phi_{2} \; (\equiv h_{1})$ and $\phi_{3} \; (\equiv h_{2})$. The corresponding canonically normalized scattering fields in the Einstein frame are then defined to be $\varphi_{1}$, $\varphi_{2}$ and $\varphi_{3}$.
Once we have determined the interactions of the canonically normalised fields, we use dimensional analysis to estimate the scale of tree-level unitarity-violation. For this purpose we introduce appropriate factors of $\tilde{E}$ to make the coefficient of the interaction terms in ${\cal L}$ dimensionless. Energy scales $\tilde{E}$ in the Einstein frame are related to the ones in the Jordan frame via $\tilde{E} = E/\Omega$, where during inflation $\Omega^2 \simeq \xi_1 \overline{\phi}_1^2/m_\text{P}^2 \approx N \gg 1$. 
Unitarity conservation implies that the matrix element for any $2\leftrightarrow 2$ scattering process should be smaller than ${\cal O}(1)$, so we can determine the scale of unitarity-violation (denoted by $\tilde{\Lambda}$) by determining the value of $\tilde{E}$ that saturates this bound.
This was demonstrated explicitly in~\cite{Lerner:2009xg}, by comparing the dimensional estimate with the exact value from the full scattering amplitude.

In the following, we consider three regimes for $\overline{\phi}_{1}$, each of which leads to a different form of the Lagrangian and the scattering amplitudes:
\begin{itemize}
 \item \emph{Regime A: Large field values}. In this regime we have $\Omega > 1$ (implying that $\xi_{1} \, \overline{\phi}_{1}^{2}/m_\text{P}^{2} > 1$) and $6 \, \xi_{1}^{2} \, \overline{\phi}_{1}^{2}/m_\text{P}^2 > 1$.
 \item \emph{Regime B: Medium field values}. In this regime we have approximately $\Omega \approx 1$ (implying that $\xi_{1} \, \overline{\phi}_{1}^{2}/m_\text{P}^{2} < 1$), but still $6 \, \xi_{1}^{2} \, \overline{\phi}_{1}^{2}/m_\text{P}^2 > 1$.
 \item \emph{Regime C: Small field values}. Finally we consider $6 \, \xi_{1}^{2} \, \overline{\phi}_{1}^{2}/m_\text{P}^2 < 1$, which in particular implies $\Omega \approx 1$.
\end{itemize}

\subsection{Regime A: Large field values}

In this case the canonically normalized fields are $\varphi_{1} = 
\sqrt{6} \, m_\text{P} \, \delta \phi_{1}/\overline{\phi}_{1}$ and $\varphi_{2,3} = \phi_{2,3}/\Omega$. The interaction leading to unitarity-violation in $\delta s \; h_{1}$ scattering is  
\be{vx1} \tilde{\mathcal{L}} \supset \frac{6 \, \xi_{1} \, \xi_{2}}{m_\text{P}^{2} \, \Omega^{4}} (\overline{\phi}_{1} + \delta \phi_{1}) \, \phi_{2} \,\tilde{g}^{\mu\nu} \, \partial_{\mu} \delta \phi_{1} \, \partial_{\nu} \phi_{2}   \;.\ee    
This results in a 3-point and a 4-point interaction. After rescaling to canonically normalized fields, the 3-point interaction is 
\be{vx2} \tilde{\mathcal{L}} \supset \frac{\sqrt{6} \, \xi_{2}}{m_\text{P}} \varphi_{2} \, \tilde{g}^{\mu\nu} \, \partial_{\mu} \delta \varphi_{1} \, \partial_{\nu} \varphi_{2} \;.\ee
This interaction can mediate $\varphi_{1} \varphi_{2} \leftrightarrow \varphi_{1} \varphi_{2}$ scattering at energy $\tilde{E}$ via $\varphi_{2}$ exchange, with a matrix element given dimensionally by $|{\cal M}| \sim  \tilde{E}^{2} \, \xi_{2}^{2}/m_\text{P}^{2}$. Unitarity is violated once $|{\cal M}| \sim 1$, therefore the unitarity-violation scale in the Einstein frame is  
\be{vx3} \tilde{\Lambda}_{12}^{(3)} \sim \frac{m_\text{P}}{\xi_{2}}   \;,\ee
where the superscript (3) denotes unitarity-violation due to the 3-point interaction.    
In the Jordan frame $\Lambda_{12}^{(3)}  = \Omega \tilde{\Lambda}_{12}^{(3)}$, where $\Omega \approx \sqrt{\xi_{1}} \;\overline{\phi}_{1}/m_\text{P}$, therefore 
\be{vx4} \Lambda_{12}^{(3)} \sim \frac{\sqrt{\xi_{1}}}{\xi_{2}} \overline{\phi}_{1}  \;.\ee
Similarly, the 4-point interaction has an Einstein frame matrix element given by $|{\cal M}| \sim \xi_{2} \, \tilde{E}^{2}/m_\text{P}^2$, therefore the 
scale of unitarity-violation in the Jordan frame is 
\be{vx5} \Lambda_{12}^{(4)} \sim \sqrt{\frac{\xi_{1}}{\xi_{2}} }  \overline{\phi}_{1}  \;.\ee 
For $\xi_{2} > 1$, this is larger than $\Lambda_{12}^{(3)}$, therefore $\Lambda_{12}^{(3)}$ is the dominant scale of unitarity-violation. In general these estimates of the unitarity-violation scales are valid if the scalars can be considered massless, which will be true if $\overline{\phi}_{1} < \Lambda_{12}^{(3)}$, i.e.\ for $\sqrt{\xi_1} > \xi_2$.

In the case of Higgs scattering $\varphi_{2} \varphi_{3} \leftrightarrow \varphi_{2} \varphi_{3}$, there is only the 4-point interaction following from 
\be{vx6}  \tilde{\mathcal{L}} \supset \frac{6 \, \xi_2 \, \xi_3}{m_\text{P}^2 \, \Omega^4} \, \phi_2 \, \phi_3 \, \tilde{g}^{\mu\nu} \, \partial_\mu \phi_2 \, \partial_\nu \phi_3   \;.\ee
As the canonically normalized Higgs fields are in general given by  $\varphi_{2,3} = \phi_{2,3}/\Omega$, the unitarity-violation scale in the Einstein frame is generally $\tilde{\Lambda}_{23} \sim m_\text{P}/\sqrt{\xi_{2} \, \xi_{3}}  \equiv m_\text{P}/\xi_{2}$ (since $\xi_{2} = \xi_{3}$ if both scalars are part of the Higgs doublet). On translating the energy to the Jordan frame, the unitarity-violation scale becomes 
\be{vx7}  \Lambda_{23} \sim \frac{\sqrt{\xi}_{1}}{\xi_{2}} \overline{\phi}_{1}   \;,\ee
which is the same expression as for $\Lambda^{(3)}_{12}$.

\subsection{Regime B: Medium field values}
 
In this case the canonically normalized fields are $\varphi_{1} = \sqrt{6} \xi_{1} \overline{\phi}_{1} \delta \phi_{1}/ m_\text{P}$ and 
$\varphi_{2} = \phi_{2}$. Since $\Omega = 1$, the energies are the same in the Einstein and Jordan frames. Using the same procedure as before, we find 
\be{vx4b} \Lambda_{12}^{(3)} \sim \frac{m_\text{P}}{\xi_{2}}  \;,\ee
and
\be{vx5b} \Lambda_{12}^{(4)} \sim \sqrt{\frac{\xi_{1}}{\xi_{2}}}  \overline{\phi}_{1}  \;.\ee
For Higgs scattering we obtain
\be{vx8} \Lambda_{23} \sim \frac{m_\text{P}}{\xi_{2}}   \;.\ee

\subsection{Regime C: Small field values}

In this case the Einstein and Jordan frames are completely equivalent. Therefore
\be{vx4c} \Lambda_{12}^{(3)} \sim \frac{m_\text{P}^{2}}{
6 \, \xi_{1} \xi_{2} \overline{\phi}_{1}}  \;,\ee
and
\be{vx5c} \Lambda_{12}^{(4)} \sim \frac{m_\text{P}}{\sqrt{\xi_{1}\xi_{2}}}  \;.\ee
For Higgs scattering we obtain
\be{vx9} \Lambda_{23} \sim \frac{m_\text{P}}{\xi_{2}}   \;.\ee

\subsection{Discussion}

In summary, for  $\xi_{s} > \xi_{h}$ the smallest (and so dominant) scale of unitarity-violation in each regime is given by:
\bea{vx10}  \textbf{A}: & \Lambda_{sh}^{(3)} \sim \Lambda_{h} \sim\frac{\sqrt{\xi_{s}}}{\xi_{h}} \overline{s} \nonumber \\
 \textbf{B}: & \Lambda_{sh}^{(3)} \sim \Lambda_{h} \sim \frac{m_\text{P}}{\xi_{h}} \nonumber  \\
 \textbf{C}: & 
\Lambda_{sh}^{(4)} \sim  \frac{m_\text{P}}{\sqrt{\xi_s \xi_h}}
\;,\eea
where $\Lambda_{sh}^{(3)} \equiv \Lambda_{12}^{(3)}$, $\Lambda_{h} \equiv \Lambda_{23}$, $\xi_{s} \equiv \xi_{1}$ and $\xi_{h} \equiv \xi_{2}$. 

In the case of Higgs Inflation, the scale of unitarity-violation is obtained as above but with $\xi_{s}$ set equal to  $\xi_{h}$, since the inflaton is now a component of $H$. During inflation $\Lambda \approx \overline{\phi}_{1}/\sqrt{\xi_{h}}$, with $\xi_{h} \sim 10^5$. Since this energy scale is less than $\overline{\phi}_{1}$, the gauge bosons become massive and only the physical Higgs scalar takes part in scattering. Since unitarity-violation requires more than one massless non-minimally coupled scalar, there is no unitarity-violation at energies less than $\overline{\phi}_{1}$. Unitarity-violation therefore occurs at $\Lambda \approx m_{W}(\overline{\phi}_{1}) \approx \overline{\phi}_1$ i.e.\ the unitarity-violation scale is essentially equal to the Higgs field value during inflation~\cite{Bezrukov:2009db,Bezrukov:2010jz}. As a result, either the new physics associated with unitarising the theory or strong coupling effects are expected to significantly modify the effective potential during inflation. It is uncertain in this case whether inflation is even possible and its predictions are unclear. 

\begin{figure}[tb]
\centering
\includegraphics[width=0.44\textwidth]{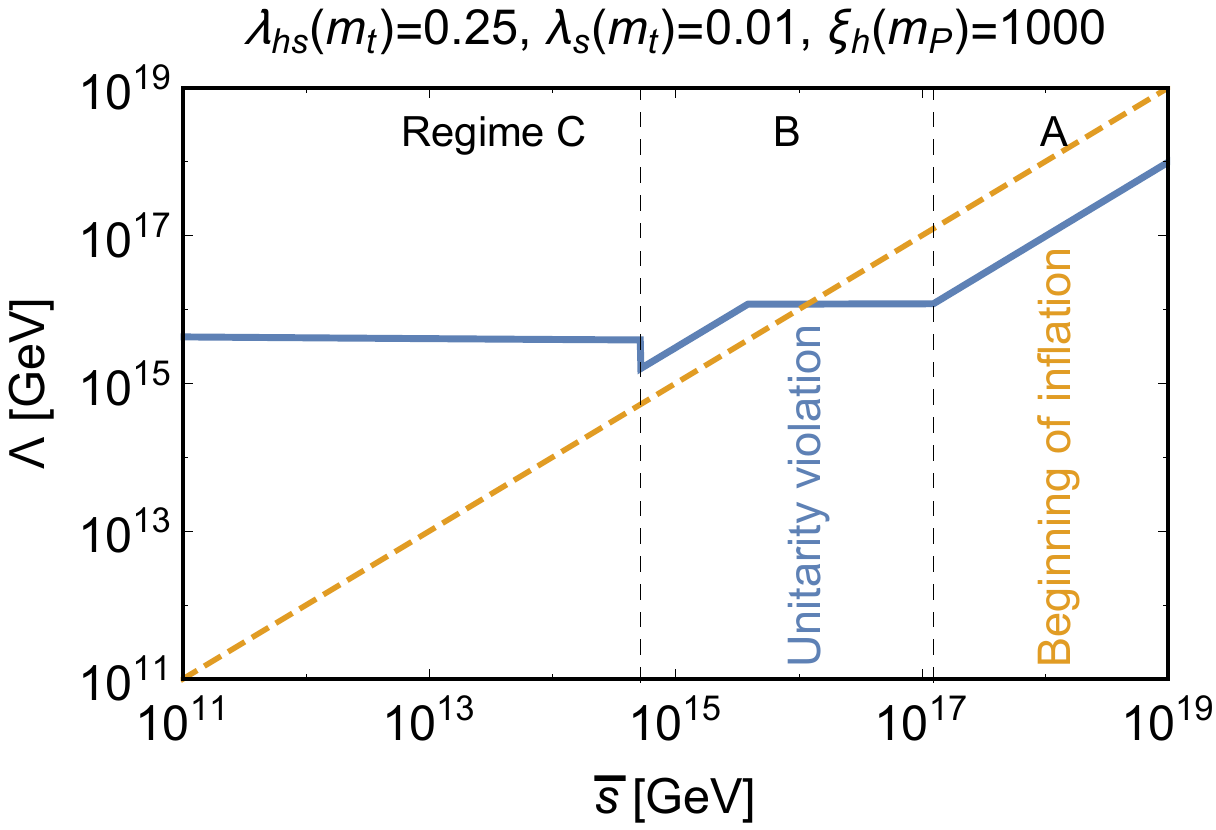}\quad
\includegraphics[width=0.44\textwidth]{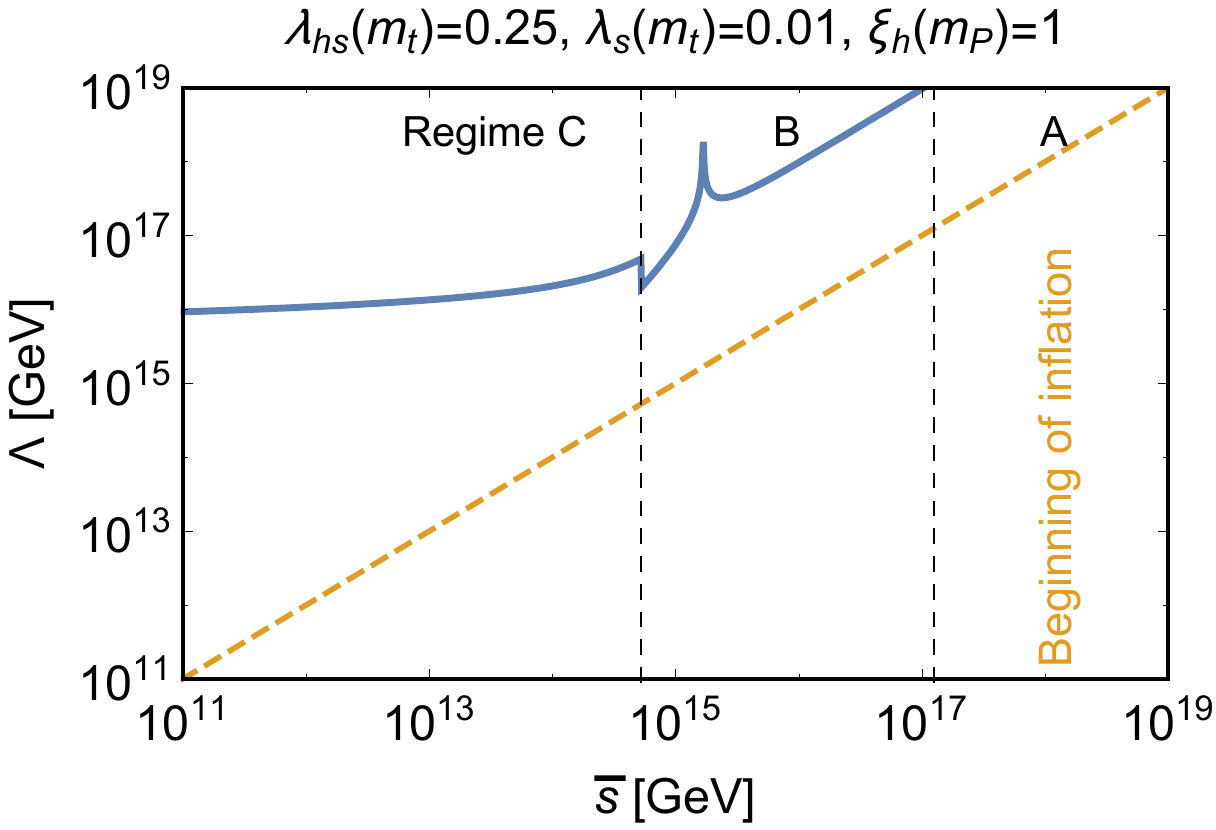}
\caption{The scale of unitarity-violation $\Lambda$ as a function of the field value $\overline{s}$ (blue) for two different choices of parameters. The orange dashed line indicates the condition $\Lambda > \overline{s}$, which must be satisfied in order to avoid unitarity-violation. In the left panel (with $\lambda_h(m_\text{P}) = 500$), unitarity is violated for $\overline{s} \gtrsim 10^{16}\:\text{GeV}$. In the right panel (with $\lambda_h(m_\text{P}) = 50$), no unitarity-violation occurs up to the scale of inflation. Note that in the right panel $\xi_h(\mu)$ runs negative for $\mu \lesssim 5 \times 10^{13}\:\text{GeV}$.}
\label{fig:running_unitarity}
\end{figure}

In S-inflation, on the other hand, it is possible to ensure that $\Lambda \gg \overline{s}$ provided $\xi_{h}$ is sufficiently small compared to $\xi_{s}$ at the scale of inflation. This is illustrated in figure~\ref{fig:running_unitarity} for $\xi_h(m_\text{P}) = 1000$ (left) and $\xi_h(m_\text{P}) = 1$ (right), taking $\lambda_{hs}(m_t) = 0.25$, $\lambda_s(m_t) = 0.01$ and $\xi_s(m_\text{P}) \approx 10^4$ as above. Both panels show the scale of unitarity-violation $\Lambda$ as a function of the field value $\overline{s}$. For $\xi_h(m_\text{P}) = 1000$ we observe that $\Lambda < \overline{s}$ for $\overline{s} \gtrsim 10^{16}\:\text{GeV}$, which is significantly smaller than the field value at the beginning of inflation. For $\xi_h(m_\text{P}) = 1$, on the other hand, $\Lambda$ always remains larger than $\overline{s}$. In this case it is reasonable to assume that new physics, in the form of additional particles with mass of order $\Lambda$ (or strong coupling effects\footnote{In unitarisation by strong coupling, $\Lambda$ is automatically field dependent and equal to the scale at which the potential is expected to change. In unitarisation by new particles, on the other hand, $\Lambda$ is only an upper bound on the masses of the new particles. Moreover the masses need not be field-dependent in order to unitarise the theory. Therefore strong coupling is more naturally compatible with the scale of inflation.}),     
will have only a small effect on the effective potential at the scale $\mu = \overline{s}$.

The right panel of figure~\ref{fig:running_unitarity} exhibits another new feature: We find $\Lambda \rightarrow \infty$ for $\overline{s} \sim 10^{15}\:\text{GeV}$. The reason is that, as already observed in figure~\ref{fig:running_high}, $\xi_h(m_\text{P})$ exhibits a strong running for $\mu < 10^{15}\:\text{GeV}$. Consequently, if we fix $\xi_h$ to a rather small value at the Planck scale, e.g.\ $\xi_h(m_\text{P}) = 1$, $\xi_h(\mu)$ will run negative at lower scales. During the transition, $\xi_h(\mu)$ will be very small and hence the scale of unitarity-violation can be very large.

It should be emphasized that the advantage of S-inflation with respect to unitarity-violation is only obtained if the singlet is a real scalar. In the case of a complex singlet, the real and imaginary parts of the scalar both have the same non-minimal coupling $\xi_{s}$. Therefore we would have $\xi_{1} = \xi_{2}$ in the above analysis and the unitarity-violation scale would become the same as in Higgs Inflation. Therefore the requirement that unitarity is not violated during inflation \emph{predicts} that the DM scalar is a real singlet scalar.  

\section{Results}
\label{results}

In this section we combine the experimental and theoretical constraints discussed above and present the viable parameter space for our model. Out of the five free parameters ($\lambda_s$, $\lambda_{hs}$, $m_s$, $\xi_s$ and $\xi_h$) we can eliminate $\lambda_{hs}$ (or $m_s$) by requiring the model to yield the observed DM abundance (see figure~\ref{fig:DM}) and $\xi_s$ by imposing the correct amplitude of the scalar power spectrum. In the following we always ensure that these two basic requirements are satisfied, and then consider additional constraints in terms of the remaining parameters $\lambda_s$, $m_s$ (or $\lambda_{hs}$) and $\xi_h$. We begin with a detailed discussion of the high-mass region and then turn to the low-mass region.

\subsection{The high-mass region}

\begin{figure}
\centering
\includegraphics[height=0.28\textheight]{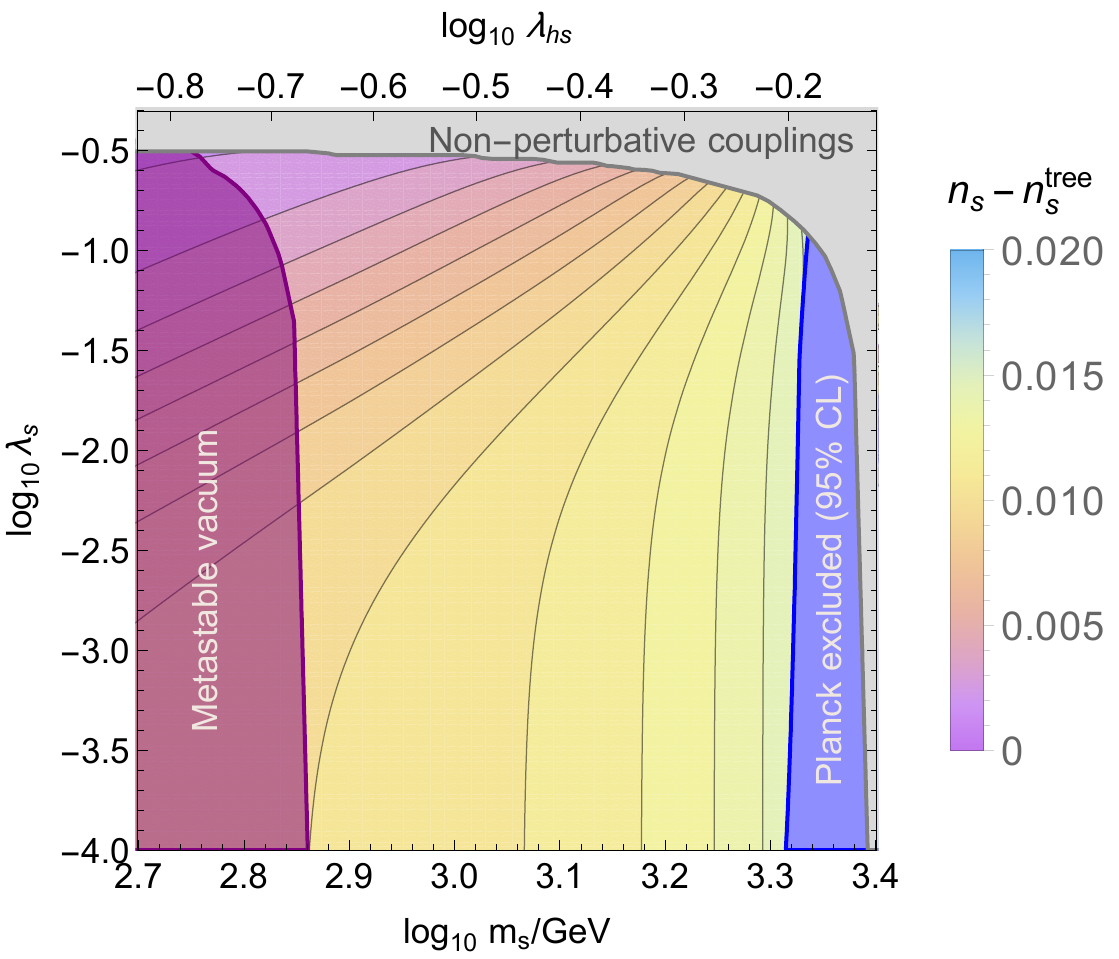}\quad
\includegraphics[height=0.28\textheight]{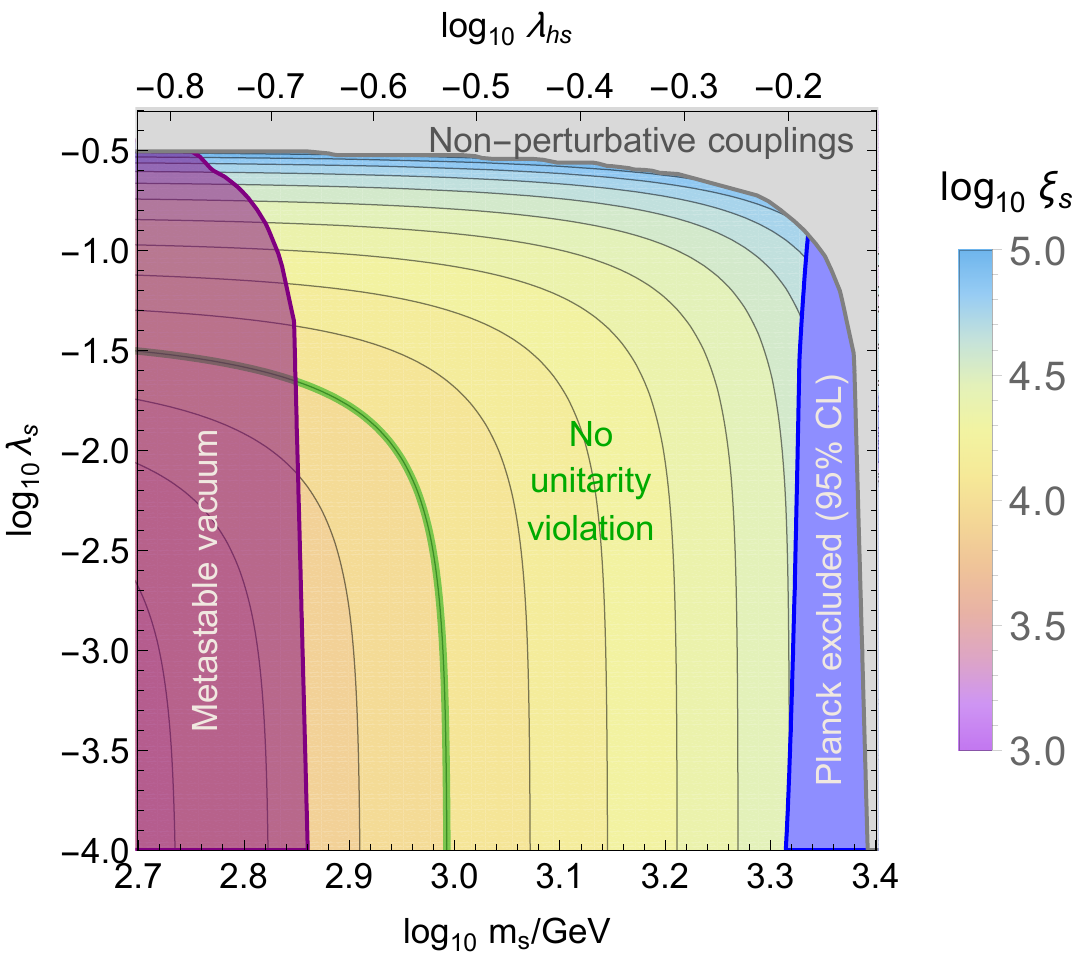}

\caption{Predictions for inflation in the high-mass region ($500\:\text{GeV} < m_s < 2500\:\text{GeV}$) as a function of $m_s$ and $\lambda_s$ for fixed $\xi_h(m_\text{P}) = 100$. Shown are the deviations from the tree-level predictions $n_s^\text{tree} = 0.965$ (left) as well as the value of $\xi_s$ at the beginning of inflation (right). For each value of $m_s$ the coupling $\lambda_{hs}$ has been fixed by the relic density requirement, as shown on the top of each panel.
The grey shaded region indicates the parameter region where couplings become non-perturbative below the scale of inflation and the purple shaded region indicates the parameter space where $\lambda_h$ runs negative below the scale of inflation, leading to a metastable vacuum.
We furthermore show the parameter region excluded by the upper bound on $n_s$ from Planck at $95\%$ CL (shaded in blue). The green line in the right panel indicates the value of $\xi_s$ where unitarity is violated at the scale of inflation.}
\label{fig:inflation_high}
\end{figure}

Let us for the moment fix $\xi_h(m_\text{P}) = 100$ and study how the predictions depend on $\lambda_s$ and $m_s$ (or, alternatively,  $\lambda_{hs}$). The left panel of figure~\ref{fig:inflation_high} shows the predicted value of $n_s$ compared to the tree-level estimate $n_s^\text{tree} = 0.965$. We find that in our model $n_s$ is always slightly larger than the tree-level estimate, but the differences are typically $\Delta n_s < 0.01$. Only for $\lambda_{hs} > 0.5$, corresponding to $m_{s} \gtrsim 2$ TeV,  do the differences grow so large that the model can be excluded by the Planck 2-$\sigma$ bound, $n_s < 0.98$. In the same parameter region we find the largest differences between the tree-level predictions of $r$ and $\alpha$ (see section~\ref{model}) and the value predicted in our model. However, we find these deviations to be negligibly small. In particular our model predicts $r < 0.01$ everywhere, i.e. the tensor-to-scalar ratio would be very difficult to observe in the near future\footnote{Next generation CMB satellites, such as PIXIE~\cite{Kogut:2011xw} and LiteBIRD~\cite{2014JLTP..176..733M}, plan to measure $r$ to an accuracy of $\delta r < 0.001$. This would be sufficient to detect the tensor-to-scalar ratio in our model.}. Figure~\ref{fig:inflation_high} also shows the parameter region excluded by the requirements that all couplings remain perturbative up to the scale of inflation (shaded in grey). This constraint requires $\lambda_s \lesssim 0.3$ for small values of $\lambda_{hs}$ and becomes more severe with increasing $\lambda_{hs}$. 

Finally, we also show the parameter region where $\lambda_{hs}$ is too small to prevent $\lambda_h$ from running to negative values in the $h$-direction (shaded in purple). While this is not fatal for the model (the electroweak vacuum remains metastable with a lifetime that is longer than the one predicted for the SM alone), this constraint may be physically significant depending upon the interpretation of the energy of the metastable state. We note, however, that the bound from metastability depends very sensitively on the assumed values of the SM parameters at the electroweak scale. Indeed, it is still possible within experimental uncertainties (at 95\% CL) that the electroweak vacuum is completely stable even in the absence of any new physics~\cite{Degrassi:2012ry}. 

For the currently preferred values of the SM parameters, we find the interesting parameter region to be $0.2 \lesssim \lambda_{hs} \lesssim 0.6$ corresponding roughly to $700\:\text{GeV} \lesssim m_s \lesssim 2\:\text{TeV}$. Very significantly, this entire range of masses and couplings can potentially be probed by XENON1T.

To study the predictions of inflation~--- and in particular the scale of unitarity-violation~--- in more detail, we show in the right panel of figure~\ref{fig:inflation_high} the value of $\xi_s$ (at the scale of inflation) required by the scalar power spectrum amplitude. We typically find values around $10^{4}$, although values as large as $10^5$ become necessary as $\lambda_{s}$ comes close to the perturbative bound. 
Since we have fixed $\xi_h(m_\text{P}) = 100$ in this plot, such large values of $\xi_s$ imply that $\sqrt{\xi_s} / \xi_h > 1$, which in turn means that the scale of unitarity-violation is larger than the field value $\bar{s}$ at the beginning of inflation.
Conversely, if both $\lambda_s$ and $\lambda_{hs}$ are small, $\xi_s$ can be significantly below $10^4$, such that unitarity is violated below the scale of inflation. The parameters for which the scale of unitarity-violation is equal to the scale of inflation is indicated by a green line.

\begin{figure}
\centering
\includegraphics[height=0.28\textheight]{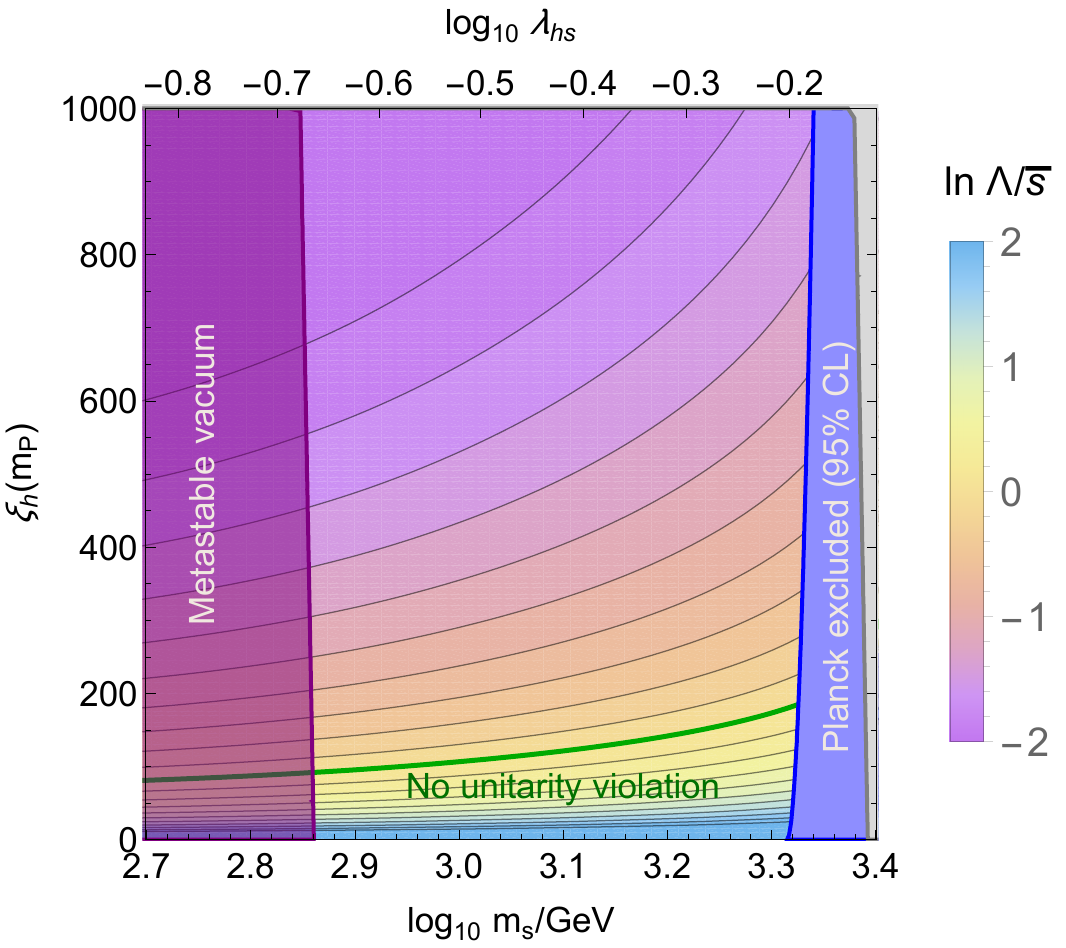}\quad
\includegraphics[height=0.28\textheight]{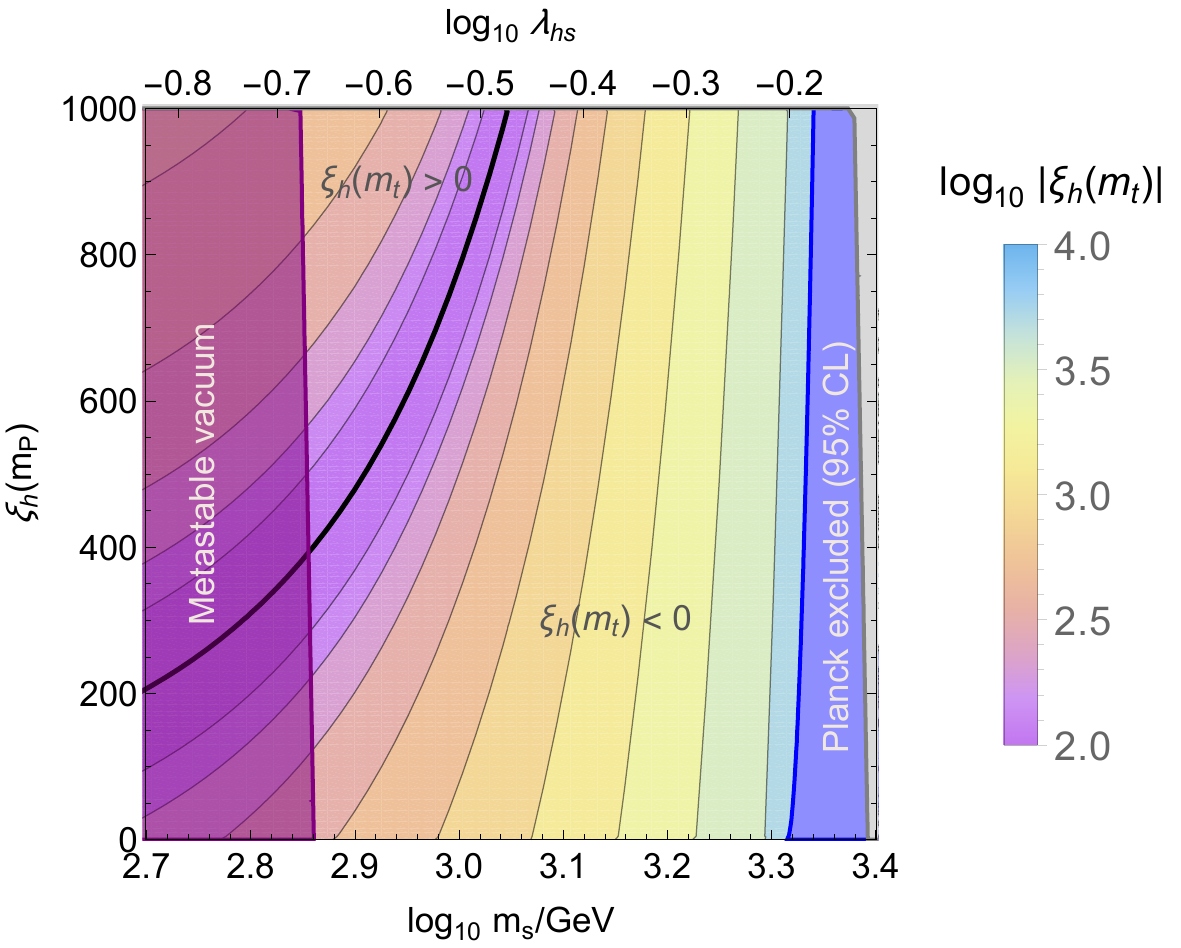}
\caption{Left: the scale of unitarity-violation $\Lambda$ compared to the field value at the beginning of inflation $\bar{s}$ as a function of $\xi_h(m_\text{P})$ and $m_s$ (or $\lambda_{hs}$) for $\lambda_s = 0.01$. In order to avoid unitarity-violation, we require $\log \Lambda / \bar{s} > 0$. Right: the corresponding value of $\xi_h$ at the electroweak scale ($\mu = m_t$). The shaded regions correspond to the same constraints as in figure~\ref{fig:inflation_high}.}
\label{fig:unitarity_high}
\end{figure}

Let us now turn to the dependence of our results on the value of  $\xi_h(m_\text{P})$. For this purpose, we fix $\lambda_s = 0.01$ and consider the effect of varying $\xi_h(m_\text{P})$ in the range $0 \leq \xi_h(m_\text{P}) \leq 1000$. We find that neither the constraint from Planck, nor the bounds from metastability and perturbativity depend strongly on $\xi_h(m_\text{P})$. Nevertheless, as discussed in section~\ref{unitarity}, $\xi_h(m_\text{P})$ does play a crucial role for the scale of unitarity-violation. We therefore show in the left panel of figure~\ref{fig:unitarity_high} the scale of unitarity-violation at the beginning of inflation divided by the field value $\bar{s}$ at the beginning of inflation. This ratio is to be larger than unity in order to avoid unitarity-violation. As indicated by the green line, this requirement implies $\xi_h(m_\text{P}) \lesssim 150$ in the parameter region of interest.

As discussed in section~\ref{RG}, small values of $\xi_h(m_\text{P})$ imply that $\xi_h(\mu)$ will run to negative values for $\mu \rightarrow m_t$. To conclude our discussion of the high-mass region we therefore show in the right panel of figure~\ref{fig:unitarity_high} the magnitude of $\xi_h(m_t)$ as a function of $m_s$ and $\xi_h(m_\text{P})$. The thick black line indicates the transition between $\xi_h(m_t) > 0$ and $\xi_h(m_t) < 0$. By comparing this plot with the one to the left, we conclude that within the parameter region that avoids unitarity-violation $\xi_h(m_t)$ necessarily becomes negative.

\subsection{The low-mass region}

\begin{figure}
\centering
\includegraphics[height=0.27\textheight]{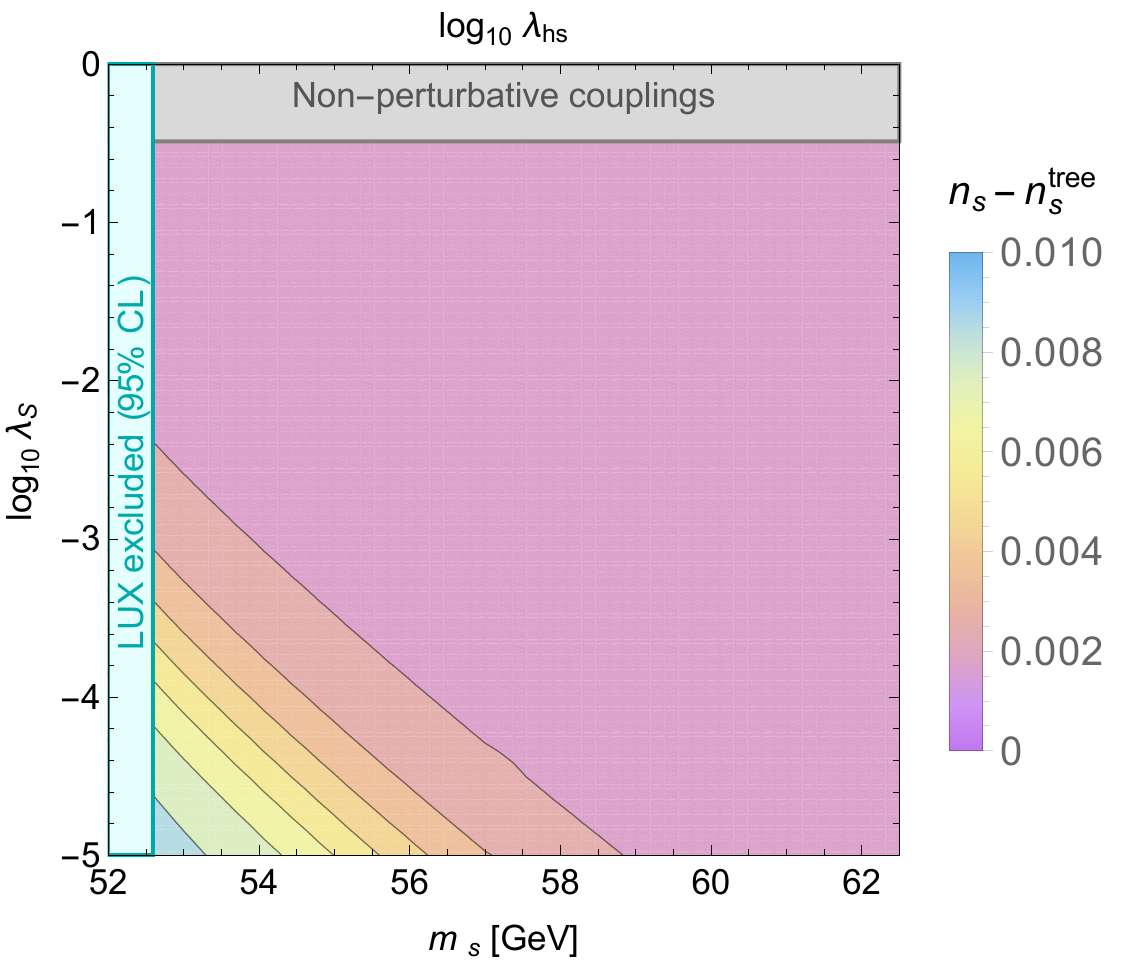}\hfill
\includegraphics[height=0.27\textheight,clip,trim={0 0 -16 0}]{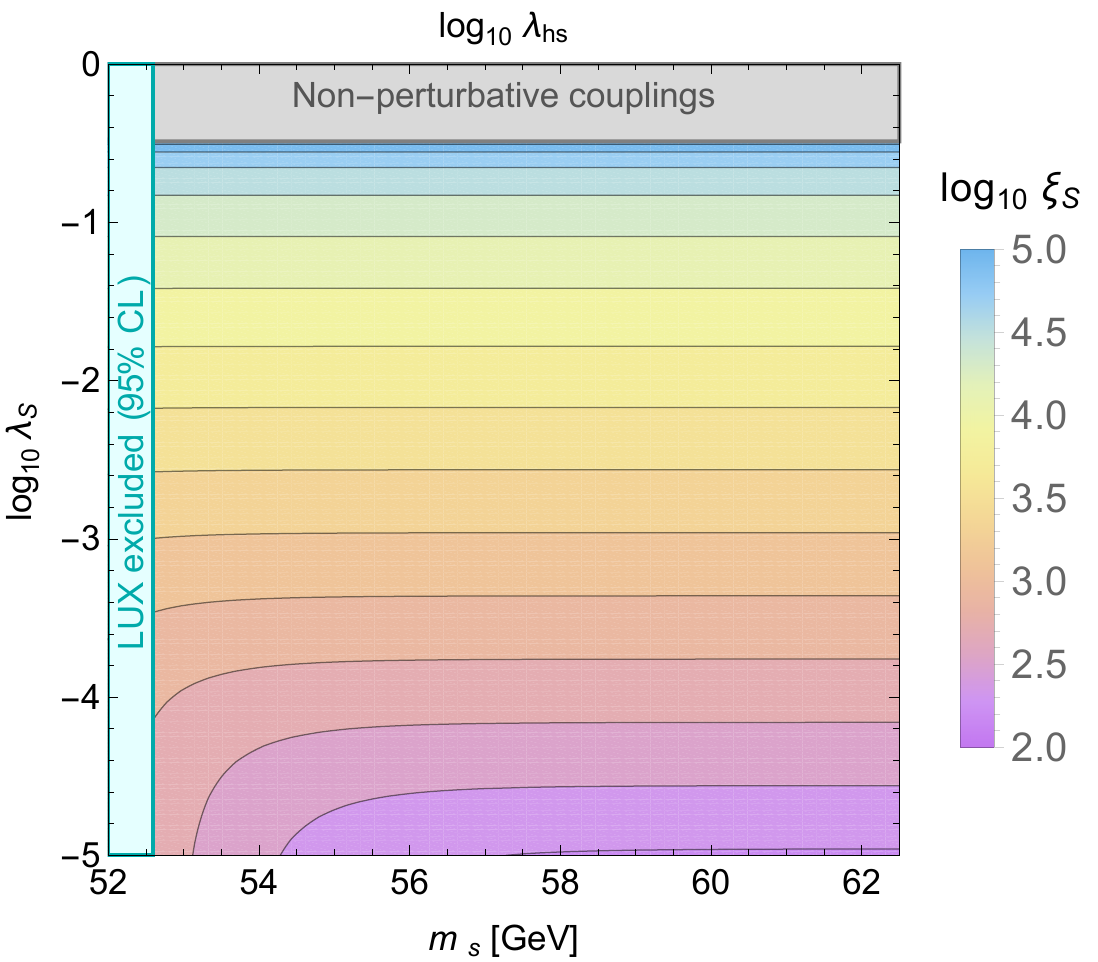}

\includegraphics[height=0.27\textheight]{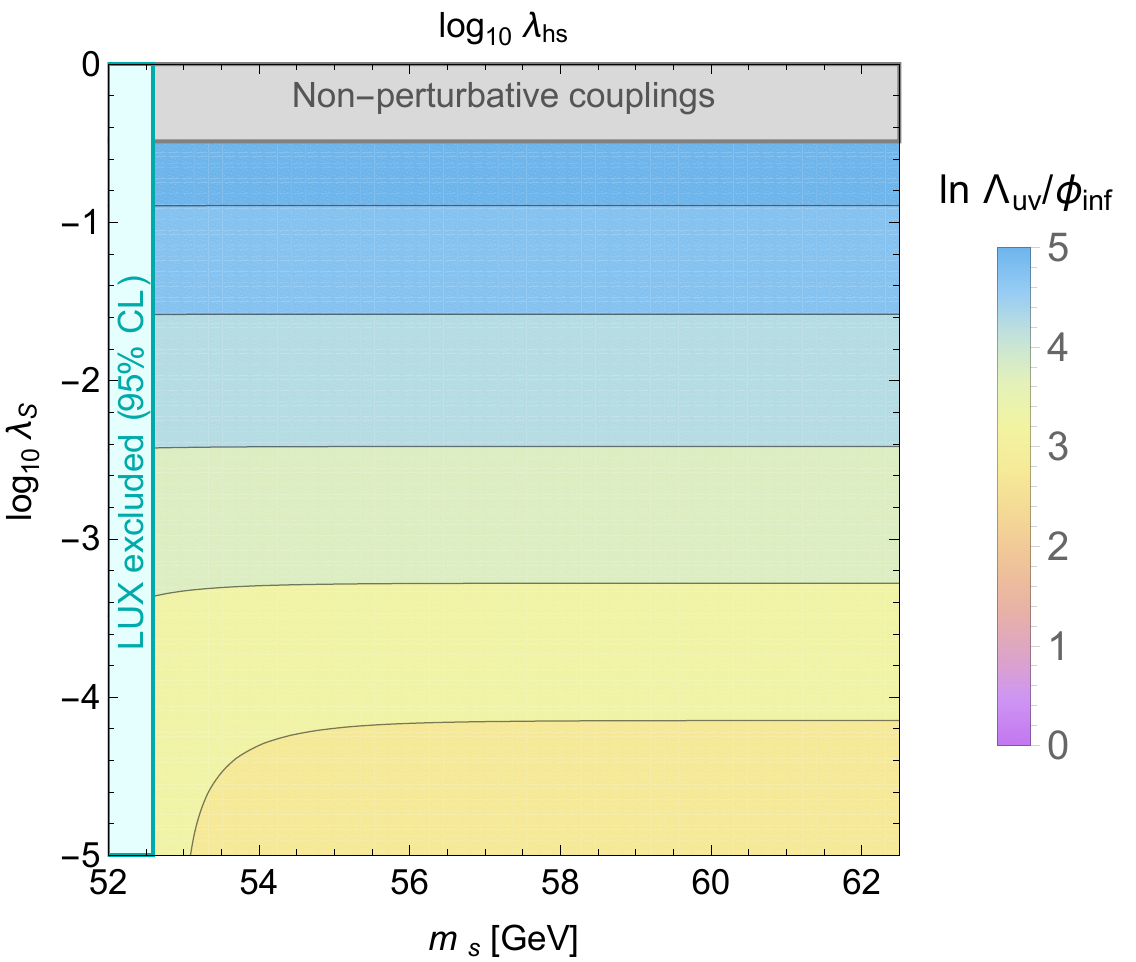}\hfill
\includegraphics[height=0.27\textheight,clip,trim={-16 0 0 0}]{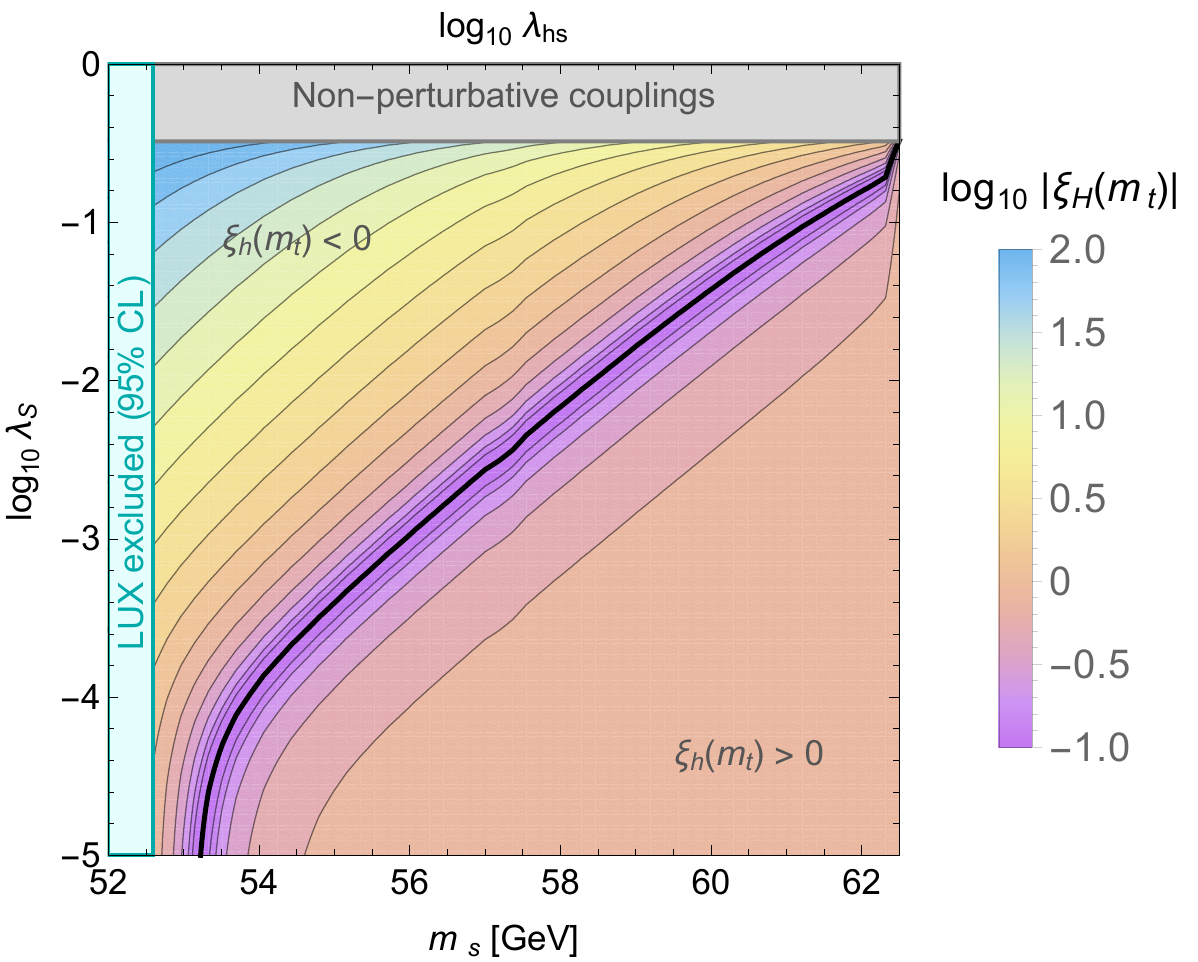}
\caption{Top row: Predictions for inflation in the low-mass region ($52.5\:\text{GeV} < m_s < 62.5\:\text{GeV}$) as a function of the couplings $\lambda_{hs}$ and $\lambda_s$ for fixed $\xi_h(m_\text{P}) = 1$. Shown are the deviations from the tree-level predictions $n_s^\text{tree} = 0.965$ (left) and the value of $\xi_s$ at the beginning of inflation (right). Bottom row: The scale of unitarity-violation $\Lambda$ compared to the field value at the beginning of inflation $s_{inf}$ (left) and the value of $\xi_h$ at the electroweak scale (right). In all panels the grey shaded region indicates the parameter region where couplings become non-perturbative below the scale of inflation and the light blue shaded region represents the 95\% CL bound from LUX. We do not show the metastability bound, since it covers the entire low-mass region.}
\label{fig:inflation_low}
\end{figure}

We study the predictions for the low-mass region in figure~\ref{fig:inflation_low}. The top row shows $\Delta n_s$ and $\xi_s$ at the beginning of inflation as a function of $\lambda_s$ and $\lambda_{hs}$ for $\xi_h(m_\text{P}) = 1$. These plots are analogous to the ones for the high-mass region in figure~\ref{fig:inflation_high}. The crucial observation is that, unless $\lambda_s \ll \lambda_{hs}$, radiative corrections to the inflationary potential are completely negligible, because any contribution proportional to $\lambda_s$ is suppressed by powers of $c_s \ll 1$ during inflation. Consequently, in most of the low-mass region, $n_s$ is identical to its tree-level value. We furthermore find that, as expected, the tensor-to-scalar ratio and the running of the spectral index are both unobservably small. Since radiative corrections play such a small role, the value of $\xi_s$ required from inflation depends almost exclusively on $\lambda_s$. As a result, it is easily possible to have $\xi_s < 1000$ during inflation 
for $\lambda_s < 10^{-3}$  and $\xi_s < 100$ for $\lambda_s < 10^{-5}$. 

Since $\xi_s$ can be much smaller in the low-mass region than in the high-mass region, it is natural to also choose a very small value for $\xi_h$. In fact, for typical values in the low-mass region $\xi_s \sim 10^3$ and $\lambda_s \sim \lambda_{hs} \sim 10^{-3}$, the loop-induced contribution to $\xi_h$ is $\Delta \xi_h < 10^{-2}$, so that it is technically natural to have $\xi_h \ll 1$. One then obtains $\sqrt{\xi_s} / \xi_h \gg 1$ and hence the scale of unitarity-violation is well above the scale of inflation. It is therefore possible without difficulty to solve the issue of unitarity-violation in the low-mass region.

This conclusion is illustrated in the bottom row of figure~\ref{fig:inflation_low}, which should be compared to figure~\ref{fig:unitarity_high} from the high-mass case, except that we keep $\xi_h(m_\text{P}) = 1$ fixed and vary $\lambda_s$ instead. The bottom-left plot clearly shows that (for our choice of $\xi_h$) the scale of 
unitarity-violation is always well above the field value at the beginning of inflation, so that the problem of unitarity-violation can easily be solved in the low-mass region. 
In addition, if the scalar couplings are sufficiently small, $\xi_h$ will have negligible running from the Planck scale down to the weak scale. It is therefore easily possible to set e.g.\ $\xi_h(m_\text{P}) = 1$ and still have $\xi_h(m_t) > 0$, as illustrated in the bottom-right panel of figure~\ref{fig:inflation_low}. It is not possible, however, to ensure at the same time that $\lambda_h$ remains positive for large field values of $h$. In other words, the electroweak vacuum is always metastable in the low-mass region (for the preferred SM parameters).

In the high-mass region we found that the coupling $\lambda_{hs}$ is bounded from below by the desire to stabilise the electroweak vacuum and from above by constraints from Planck and the requirement of perturbativity. In the low-mass region, on the other hand, we obtain an upper bound on $\lambda_{hs}$ from LUX and a lower bound on $\lambda_{hs}$ from the relic density requirement. Compared to the high-mass region, the allowed range of couplings in the low-mass region is much larger and therefore much harder to probe in direct detection experiments. If indeed $m_s$ is very close to $m_h / 2$, and $\lambda_s,\,\lambda_{hs} < 10^{-3}$, it will be a great challenge to test the model predictions with cosmological or particle physics measurements.

\section{Conclusions}\label{conclusions}

The origin of inflation and the nature of DM are two of the fundamental questions of cosmology. In the present work, we have revisited the possibility that both issues are unified by having a common explanation in terms of a real gauge singlet scalar, which is one of the simplest possible extensions of the SM. Considering the most recent experimental constraints for this model from direct detection experiments, the LHC and Planck, we have shown that large regions of parameter space remain viable. Furthermore, we find that in parts of the parameter space the scalar singlet can stabilise the electroweak vacuum all the way up to the Planck scale, while at the same time avoiding the problem of unitarity-violation present in conventional models of Higgs inflation.

The scalar singlet can efficiently pair-annihilate into SM particles via the Higgs portal, so that it is straight-forward in this model to reproduce the observed DM relic abundance via thermal freeze-out. We find two distinct mass regions where the model is consistent with experimental constraints from LUX, LHC searches for invisible Higgs decays and Fermi-LAT:
the low-mass region, $53\:\text{GeV} \lesssim m_s \lesssim 62.4\:\text{GeV}$, where DM annihilation via Higgs exchange receives a resonant enhancement, and the high-mass region, $m_s \gtrsim 93\:\text{GeV}$, where a large number of annihilation channels are allowed. 

In both mass regions it is possible without problems to fix the non-minimal couplings $\xi_s$ and $\xi_h$ in such a way that inflation proceeds in agreement with all present constraints. 
In particular, the tensor-to-scalar ratio and the running of the spectral index are expected to be unobservably small. On the other hand, radiative corrections to the spectral index typically lead to a value of $n_s$ slightly larger than the classical estimate, i.e.\ $n_s > 0.965$. This effect is largest for large values of $m_s$ and $\lambda_{hs}$ and current Planck constraints already require $m_s \lesssim 2 \:\text{TeV}$. The entire high-mass region compatible with Planck constraints will therefore be tested by XENON1T, which can constrain gauge singlet scalar DM up to $m_s \sim 4\:\text{TeV}$.

In the high-mass region, the value of $\xi_s$ required to obtain a sufficiently flat potential during inflation is typically $\xi_s \sim 10^4\text{--}10^5$. In spite of such a large non-minimal coupling, it is possible to have unitarity-conservation during inflation, in the sense that the scale of unitarity-violation can be much larger than the inflaton field. The reason is that only $\xi_s$ needs to be large in order to reproduce the observed density perturbation, while the Higgs non-minimal coupling $\xi_h$ can be arbitrarily small. In the limit $\xi_{h} \rightarrow 0$ there will be only one non-minimally coupled scalar field and therefore no unitarity-violation,  provided that the inflaton, and so the DM particle, is a \emph{real} scalar.

We find that at large singlet field values $\bar{s}$ the scale of unitarity-violation is given by $\Lambda \sim \bar{s} \sqrt{\xi_s}/\xi_h$. If the non-minimal couplings satisfy $\xi_{s}(m_\text{P}) \gg \xi_{h}(m_\text{P})$ at the Planck scale, it is possible for the unitarity-violation scale during inflation to be orders of magnitude larger than $\bar{s}$. Such a hierarchy of couplings is stable under radiative corrections and consistent with the assumption that inflation proceeds along the $s$-direction. 
Furthermore, in the low-mass region $\lambda_{hs}$ and $\lambda_s$ can be so small that $\xi_s \sim 10^2 \text{--} 10^3$ is sufficient to obtain a flat enough potential. 

We conclude that it is possible for the inflaton potential in S-inflation to be safe from new physics or strong-coupling effects associated with the unitarity-violation scale. This contrasts with the case of Higgs Inflation, where unitarity is always violated at the scale of the inflaton. 

Another interesting observation is that if the singlet mass and the coupling $\lambda_{hs}$ are sufficiently large (roughly $m_s \gtrsim 1\:\text{TeV}$ and $\lambda_{hs} \gtrsim 0.3$), the presence of the additional scalar singlet stabilises the electroweak vacuum, because the additional contribution to $\beta_{\lambda_h}$ prevents the quartic Higgs coupling from running to negative values. This observation becomes important if a metastable electroweak vacuum is physically disfavoured, for example if the potential energy relative to the absolute minimum defines an observable vacuum energy. 

Given how tightly many models for DM are constrained by direct detection and LHC searches and how strong recent bounds on models for inflation have become, it is quite remarkable that one of the simplest models addressing both problems still has a large allowed parameter space. Nevertheless, the model is highly predictive. In particular, if the DM scalar is also the inflaton and unitarity is conserved during inflation, then DM is predicted to be a real scalar. Direct detection experiments will soon reach the sensitivity necessary to probe the entire parameter space relevant for phenomenology, with the exception of a small window in $m_{s}$ close to the Higgs resonance. The next few years will therefore likely tell us whether indeed a singlet scalar extension of the SM can solve two of the central problems of particle physics and cosmology.

\acknowledgments

We would like to thank Rose Lerner for her contribution during the early stages of this project. We are grateful to Kyle Allison, Ido Ben-Dayan, Andreas Goudelis, Huayong Han, Thomas Konstandin, Kai Schmidt-Hoberg, Pat Scott and Christoph Weniger for helpful discussions, and to Guillermo Ballesteros for carefully reading the manuscript and providing a number of useful comments. The work of FK was supported by the German Science Foundation (DFG) under the Collaborative Research Center (SFB) 676 Particles, Strings and the Early Universe. The work of JMcD was partly supported by the Lancaster-Manchester-Sheffield Consortium for Fundamental Physics under STFC grant ST/J000418/1. FK would like to thank the Instituto de Fisica Teorica (IFT UAM-CSIC) in Madrid for its support via the Centro de Excelencia Severo Ochoa Program under Grant SEV-2012-0249, during the Program ``Identification of Dark Matter with a Cross-Disciplinary Approach'' where part of this work was carried out.

\renewcommand{\theequation}{A-\arabic{equation}}  \setcounter{equation}{0} 
\section*{Appendix: RG Equations} 

The RG equations for the scalar couplings can be obtained using the techniques detailed in~\cite{Machacek:1983tz,Machacek:1983fi,Machacek:1984zw}, as in~\cite{Lerner:2009xg}. The one-loop $\beta$-functions for the scalar couplings are
\begin{align} 16\pi^2 \, \beta_{\lh}^{(1)}  & = - 6 \, y_t^4 + \frac{3}{8}\left(2 \, g^4 + \left(g^2+g'^2\right)^2\right) + \left(-9 \, g^2 -3 \, g'^2 + 12 \, y_t^2\right)\lh \nonumber \\ & \quad +\left(18 \, c_h^2 + 6\right)\lh^2+  \frac{1}{2} \, c_s^2 \, \lhs^2
\;, \\
&\nonumber\\
16\pi^2 \, \beta_{\lhs}^{(1)} & = 4 \, c_h \, c_s \, \lhs^2 + 6\left(c_h^2+1\right)\lh \, \lhs  - \frac{3}{2}\left(3 \, g^2 + g'^2\right)\lhs \nonumber \\ 
& \quad  + 6 \, y_t^2 \, \lhs  + 6 \, c_s^2 \, \ls \, \lhs \; ,\\
&\nonumber\\
16\pi^2 \, \beta_{\ls}^{(1)} & = \frac{1}{2}(c_h^2 + 3)\lhs^2  +18 \, c_s^2 \, \ls^2
\;.
\end{align}
The propagator suppression factors are given by 
\bea{supeq}
c_{\phi} = \frac{1 + \frac{\xi_{\phi} \, {\phi}^2}{m_\text{P}^2}}{1 + (6 \, \xi_{\phi} + 1)\frac{\xi_{\phi} \, {\phi}^2}{m_\text{P}^2}} \; ,
\eea
where $\phi$ is $s$ or $h$.

The RG equations for the non-minimal coupling can be derived following~\cite{Buchbinder:1992rb}, as in~\cite{Lerner:2009xg} (see also~\cite{Clark:2009dc}). One obtains
\begin{align}
16\pi^2 \, \frac{\mathrm{d}\xi_s}{\mathrm{d}t} & =  \left(3+c_h\right)\lhs\left(\xi_h+\frac{1}{6}\right)
+\left(\xi_s+\frac{1}{6}\right)6 \, c_s \, \ls \; , \\
& \nonumber \\
16\pi^2 \, \frac{\mathrm{d}\xi_h}{\mathrm{d}t} & = \left(\left(6+6\,c_h\right)\lh + 6 \, y_t^2 - \frac{3}{2}(3 \, g^2 + g'^2)\right)\left(\xi_h+\frac{1}{6}\right)+ \left(\xi_s+\frac{1}{6}\right)c_s \, \lhs
\;.
\end{align}

\providecommand{\href}[2]{#2}\begingroup\raggedright\endgroup

\end{document}